\documentclass[a4paper,twocolumn,10pt,unpublished]{quantumarticle}
\pdfoutput=1
\usepackage[utf8x]{inputenc}
\usepackage[english]{babel}
\usepackage[T1]{fontenc}
\usepackage{amsmath,amssymb,amsthm}
\usepackage{booktabs}
\usepackage{graphicx}
\usepackage{xcolor}
\usepackage{wrapfig}
\usepackage{bbm}
\usepackage{listings}
\definecolor{codegreen}{rgb}{0,0.6,0}
\definecolor{codegray}{rgb}{0.5,0.5,0.5}
\definecolor{codepurple}{rgb}{0.58,0,0.82}
\definecolor{codered}{rgb}{0.796,0.235,0.2}
\definecolor{backcolour}{rgb}{0.95,0.95,0.92}
\definecolor{dodgerblue4}{rgb}{0.06, 0.31, 0.55}

\lstdefinelanguage{Julia}{
    keywords=[1]{abstract, break, catch, const, continue, do, else, elseif, begin, end, export, false, for, function, if, import, let, local, macro, module, quote, return, struct, true, try, using, while, where},
    keywords=[2]{Bool, Char, Dict, Float64, Int, String, Array, AbstractArray, Vector, Set, Tuple, NamedTuple, QuantumObject, QuantumObjectEvolution},
    keywords=[3]{dummy_start, versioninfo, Qobj, QobjEvo, tensor, basis, fock, destroy, qeye, sigmax, sigmay, sigmaz, sigmam, sigmap, range, sesolve, mesolve, mcsolve, ssesolve, smesolve, sqrt, sin, cos, steadystate, steadystate_fourier, dfd_mesolve, dsf_mesolve, coherent, sum, size, dropdims, H_dsf, c_ops_dsf, e_ops_dsf, cu, CuVector, wigner, normalize, addprocs, SlurmManager, set_num_threads, Lattice, mapreduce, multisite_operator, DissipativeIsing, Val, ntuple, EnsembleSplitThreads, rmprocs, workers, gradient, my_f_mesolve, real, expect, BacksolveAdjoint, EnzymeVJP, liouvillian, exp, conj, coef_a, coef_ac, dummy_end},
    sensitive=true,  % Julia is case-sensitive.
    morecomment=[l]\#,%
    morecomment=[n]{\#=}{=\#},%
    morestring=[s]{"}{"},%
    literate={
        {\\}{{{\color{codered}\lstum@backslash}}}{1} {\{}{{{\color{codered}\{}}}{1}
        {\}}{{{\color{codered}\}}}}{1} % {!}{{{\color{codered}!}}}{1}
        {\%}{{{\color{codered}\%}}}{1} {&}{{{\color{codered}\&}}}{1}
        {+}{{{\color{codered}+}}}{1}
        {*}{{{\color{codered}*}}}{1}
        {/}{{{\color{codered}/}}}{1}
        {'}{{{\color{codered}\textquotesingle}}}{1}
        {=}{{{\color{codered}=}}}{1}
        {<}{{{\color{codered}<}}}{1}
        {=>}{{{\color{codered}=>}}}{1} {|>}{{{\color{codered}|>}}}{1}
        {==}{{{\color{codered}==}}}{1}
        {>}{{{\color{codered}>}}}{1} {?}{{{\color{codered}?}}}{1}
        {julia>}{{{\color{codered}julia>}}}{1}
        {^}{{{\color{codered}\textasciicircum}}}{1} {|}{{{\color{codered}|}}}{1}
        {~}{{{\color{codered}\textasciitilde{}}}}{1}
        {×}{{{\color{codered}$\times$}}}{1} {⋅}{{{\color{codered}$\cdot$}}}{1}
        {π}{{$\pi$}}1
        {ω}{{$\omega$}}1
        {α}{{$\alpha$}}1
        {β}{{$\beta$}}1
        {σ}{{$\sigma$}}1
        {ψ}{{$\psi$}}1
        {κ}{{$\kappa$}}1
        {γ}{{$\gamma$}}1
        {κ}{{$\kappa$}}1
        {Δ}{{$\Delta$}}1
        {φ}{{$\phi$}}1
        {ε}{{$\varepsilon$}}1
        {∘}{{$\circ$}}1
    }
}

\lstdefinestyle{mainstyle}{
    language=Julia,
    backgroundcolor=\color{backcolour},
    commentstyle=\color{gray},    % Style for comments.
    keywordstyle=\color{blue}\bfseries,  % Style for Group 1 keywords.
    keywordstyle=[2]\color{teal},  % Style for Group 2 keywords.
    keywordstyle=[3]\color{dodgerblue4},  % Style for Group 3 keywords.
    stringstyle=\color{codepurple},
    basicstyle=\ttfamily\footnotesize,
    columns=fullflexible,
    upquote=true,
    breakatwhitespace=false,
    breaklines=true,
    captionpos=b,
    keepspaces=true,
    showspaces=false,
    showstringspaces=false,
    showtabs=false,
    tabsize=2,
}

\usepackage[numbers,sort&compress]{natbib}
\definecolor{darkred}{rgb}{0.6,0.05,0.05}
\definecolor{darkgreen}{rgb}{0.05,0.6,0.05}
\definecolor{darkblue}{rgb}{0.05,0.05,0.6}
\usepackage[colorlinks]{hyperref}
\hypersetup{colorlinks,
linkcolor=darkred,
filecolor=darkgreen,
urlcolor=darkblue,
citecolor=darkgreen}

\usepackage{microtype}
\usepackage{physics}
\usepackage{tikz}
\usetikzlibrary{arrows.meta,positioning,calc,shapes.geometric,fit,backgrounds}

\graphicspath{{./figures/}}

\newcommand{\Heff}{\hat{H}_{\mathrm{eff}}}
\newcommand{\diag}{\operatorname{diag}}
\newcommand{\Var}{\operatorname{Var}}
\newcommand{\code}[1]{\texttt{#1}}

\newcommand{\reportfig}[2]{%
  \IfFileExists{./figures/#1.png}%
    {\includegraphics[width=#2]{#1.png}}%
    {\IfFileExists{./figures/#1.pdf}%
      {\includegraphics[width=#2]{#1.pdf}}%
      {\fbox{\parbox[c][0.42\linewidth][c]{#2}{\centering\small
        \textcolor{red}{expected figure file}\par
        \code{\detokenize{#1}.\{png,pdf\}}\par
        \textcolor{red}{not present in this build}\par}}}}%
}

\title{Diagonal, Switching, and Locally Optimal Unraveling for efficient quantum trajectories in metastable open quantum systems}

\author[alicebob,epfl1,epfl2]{Pietro Pacchioni}
\orcid{0000-0001-7496-3836}
\email{pietro.pacchioni@usherbrooke.ca}
\thanks{Present address: Institut quantique \&
D\'epartement de Physique, Universit\'e de Sherbrooke,
Sherbrooke, Qu\'ebec J1K 2R1, Canada.}

\author[epfl1,epfl2]{Alberto Mercurio}
\orcid{0000-0001-7814-1936}
\email{alberto.mercurio@epfl.ch}

\author[alicebob]{Patrick Winkel}
\orcid{0000-0001-6457-6971}
\email{patrick.winkel@alice-bob.com}

\author[alicebob]{Fabrizio Minganti}
\orcid{0000-0003-4850-1130}
\email{fabrizio.minganti@gmail.com}

\affiliation[alicebob]{Alice \& Bob, 53 boulevard du G\'en\'eral Martial Valin,
75015 Paris, France}

\affiliation[epfl1]{Institute of Physics, Ecole Polytechnique Fédérale de Lausanne (EPFL), CH-1015 Lausanne, Switzerland}

\affiliation[epfl2]{Center for Quantum Science and Engineering, Ecole Polytechnique Fédérale de Lausanne (EPFL), CH-1015 Lausanne, Switzerland}

\begin{document}

\begin{abstract}
Open quantum systems can display metastability, with multiple long-lived states coexisting. 
Examples include the bistability of bright and dim manifolds in nonlinear driven resonators, or the long-lived coherent states of opposite phases in parametrically-driven systems, the building block of cat-state-based quantum computers.  
Metastability often emerges at large system sizes where a numerical evolution of the density matrix is impractical.
Quantum trajectories mitigate this issue by evolving wavefunctions rather than density matrices. 
Metastable states then emerge as attractors towards which the system rapidly relaxes until a rare switching event brings it to another attractor.
Simulating these events is still very costly because it requires resolving the fast dynamics within each attractor manifold while integrating over timescales long enough to accumulate sufficient switching statistics.
To tackle this problem, we introduce the Diagonal, Switching, and Locally Optimal Unraveling (\mbox{DiSLOU}) method, which combines three optimization layers:
(i) Switching among locally optimal unravelings that minimize the number of quantum jumps in each metastable region.
(ii) Diagonal propagation: the no-jump evolution is accelerated by diagonalizing the non-Hermitian effective Hamiltonian once per attractor. 
(iii) Local reduced eigenspaces: the no-jump state can be faithfully propagated in a truncated subspace of only slowly decaying eigenmodes.
We benchmark \mbox{DiSLOU} on the driven-dissipative Kerr resonator, finding no measurable loss of accuracy against standard quantum trajectories, with the largest performance improvement observed in the bistable regime where standard trajectories are most expensive.
We then apply the method to the more challenging problem of accurately simulating rare bit-flip events in the cat manifold emerging in a two-mode dissipatively stabilized parametric oscillator, obtaining a performance improvement of over one order of magnitude. An implementation of \mbox{DiSLOU} in Julia is available as an open-source package~\cite{DiSLOU_git}.
\end{abstract}

\maketitle

\section{Introduction}

In many driven open quantum systems, an initial state evolves towards its steady state on timescales dictated by the excitation lifetimes~\cite{carusotto2013quantum,hartmann2016quantum}. 
When the drive competes with other terms (dissipation, detuning, high-order nonlinearities, etc.) in determining the steady-state of the system, exponentially slow timescales can emerge and states can become \emph{metastable}. 
The processes occurring during this slow decay to the steady state often carry the physics of interest, e.g., the metastable switching near dissipative phase transitions~\cite{macieszczak2016,fitzpatrick2017,rodriguez2017,minganti2018, fink2018,Beaulieu2025}, or the loss of quantum information in dissipatively stabilized cat qubits~\cite{mirrahimi2014,lescanne2020,berdou2023,reglade2024}.
Studying these systems is, however, numerically challenging, since one must resolve the fast local dynamics that are present at all times while also integrating over the much longer timescales associated with metastability. 
For a bosonic Hilbert space of dimension $N$, an $N\times N$ density matrix requires $O(N^3)$ operations per time step to be evolved\footnote{\label{fn:sparse}$O(N^2)$ for sparse matrices}, which becomes prohibitively costly for large $N$~\cite{breuer1997stochastic,benenti2007principles}.
Corner-space, low-rank, and moving-basis approximations aim at reducing this cost by projecting the dynamics onto a relevant subspace~\cite{schack1995,schack1996,chamberland2022,Schlegel2023,finazzi2015,lebris2015}, while Krylov-subspace approaches target the slow Liouvillian sector directly, either through spectral methods or by constructing it from comparatively short time evolutions~\cite{am-shallem2015,can2019,minganti2022arnoldi,nandy2025}.

The quantum trajectories (also known as Monte Carlo wavefunctions) method~\cite{diosi1986,dalibard1992,dum1992,molmer1993,carmichael1993,plenio1998,daley2014} makes the problem more tractable by replacing the evolution of the $N\times N$ density matrix with stochastic wavefunction evolutions in an $N$-dimensional Hilbert space. A single sampled wavefunction evolution is called a quantum trajectory and evolves under an effective non-Hermitian Hamiltonian, interrupted at random times by the instantaneous action of jump operators (quantum jumps).
For a given open quantum system, the jump operators and non-Hermitian Hamiltonian are not uniquely determined.
Therefore, a given choice for the stochastic representation of the master equation is called an \emph{unraveling}.

In the presence of metastability, an algorithm must propagate the trajectory over long timescales.
As quantum jumps occur, interrupting the action of the non-Hermitian Hamiltonian, the dynamics display rapid oscillations.
Propagating them over very long times, together with processing the many jumps, is the dominant numerical cost.
The individual issues that make metastability hard to simulate have been addressed separately in the literature, but not in combination.
For example, event-driven approaches such as the quantum Gillespie method avoid integration over small time steps by directly sampling the time and type of the next jump~\cite{radaelli2024}, but still process every jump in the trajectory. A separate line of work targets the unraveling itself. Shifting or mixing the jump operators yields equivalent representations of the same Lindblad dynamics, altering individual trajectories while preserving their ensemble average~\cite{breslin1995,wiseman2009}. Such choices may depend on previous jumps~\cite{lucas2013,lucas2014},
specify the between-jump evolution~\cite{chruscinski2022},
reduce the sampling variance of a chosen observable~\cite{cao2026},
or optimize the unraveling to reduce trajectory entanglement and the cost of
classical simulation~\cite{vovk2022,cheng2023,kolodrubetz2023,
chen2024unraveling,vovk2024,daraban2025,cichy2026,fioroni2026}. Biased ensembles instead reweight whole trajectories to favor rare events~\cite{garrahan2010,carollo2018}, but without reducing the cost of each individual trajectory.

None of these approaches alone makes metastable switching-rate calculations efficient, and their advantages are not straightforward to combine.
We thus introduce \mbox{DiSLOU}, an effective method to significantly reduce the numerical cost of simulating quantum trajectories in metastable systems, built from three composable layers:
\begin{itemize}
    \item Layer~I, \emph{locally optimal unraveling}. The number of quantum jumps is a property of the unraveling.
    As the unraveling can be changed along a quantum trajectory, we dynamically choose the one that minimizes the jump rate for states in the vicinity of each metastable attractor. This results in longer no-jump evolutions, that are efficiently simulated by Layer~II.
    \item Layer~II, \emph{diagonal evolution}. We diagonalize the non-Hermitian Hamiltonian associated with each optimal unraveling chosen for Layer~I, so that the no-jump evolution becomes an inexpensive operation. 
    \item Layer~III, \emph{local reduced eigenspaces}. We exploit the structure of the non-Hermitian Hamiltonian of each optimal unraveling to truncate the Hilbert-space dimension, with a fallback procedure that returns to the full space when required --- for example, to describe the switching between metastable attractors.
\end{itemize}

We test the three layers on the driven-dissipative Kerr resonator.  
\mbox{DiSLOU} quantum trajectories agree, within statistical uncertainty, with master-equation evolution, but with a significant speedup relative to standard quantum trajectories. 
As a more demanding application, we study a two-mode parametrically coupled driven-dissipative system, known for hosting Schr\"odinger cat states. 
Here, we also examine the effects induced by a strongly coupled two-level system (TLS) \cite{simmonds2004,shalibo2010,sarabi2015,muller2019tls}.
Also in this case \mbox{DiSLOU} outperforms standard quantum trajectories without meaningful deviations in the resulting statistics, even when performing advanced tasks such as estimating the decay rate of quantum information stored in the system.

An implementation of \mbox{DiSLOU} in Julia, built on top of the quantum-trajectory solver of \code{QuantumToolbox.jl}~\cite{mercurio2025}, is available as an open source package~\cite{DiSLOU_git}.

The remainder of the paper is organized as follows. Section~\ref{sec:unraveling_geometry} reviews quantum trajectories and their numerical cost, Sec.~\ref{sec:method} details the three layers of \mbox{DiSLOU}, and Secs.~\ref{sec:kerr} and~\ref{sec:cat} benchmark the method on the driven-dissipative Kerr resonator and two-mode dissipative cat systems, respectively. We conclude with a discussion of limitations and perspectives in Sec.~\ref{sec:conclusion}.

\section{Quantum trajectories and their numerical cost}
\label{sec:unraveling_geometry}

We consider a finite-dimensional, time-independent Lindblad master equation for the reduced density matrix $\hat\rho$ of a quantum system, reading ($\hbar=1$)
\begin{align}
\dot{\hat\rho}
  &=-i[\hat H,\hat\rho]+\sum_{\mu=1}^{N_{\rm c}}\mathcal D[\hat C_\mu]\hat\rho,
  \nonumber\\
  \mathcal D[\hat C_\mu]\hat\rho
  &=\hat C_\mu\hat\rho \hat C_\mu^\dagger-\frac12\{\hat C_\mu^\dagger \hat C_\mu,\hat\rho\}.
  \label{eq:lindblad}
\end{align}
Here $\hat H$ is the Hamiltonian of the system and $\mathcal D[\hat C_\mu]$ is the dissipation superoperator for the jump operator $\hat C_\mu$.
These $N_{\rm c}$ operators describe the incoherent interaction of the system with its environment.

\subsection{Standard unraveling}
\label{sec:standard_unraveling}

The evolution of this system can be recast in terms of a stochastic wavefunction $\ket{\psi(t)}$ \cite{dalibard1992,dum1992,molmer1993,carmichael1993,plenio1998,daley2014}, known as a quantum trajectory. 
The density matrix is the ensemble average over many quantum trajectories, and Eq.~\eqref{eq:lindblad} follows from $\mathbb E[\ketbra{\psi(t)}{\psi(t)}]=\hat\rho(t)$.
For the quantum-jump formulation considered here, the system evolves as follows. 
At each time step $\dd t$, a random number $u\in [0,1)$ is drawn and compared with the jump probability $\dd p=\dd t\sum_{\mu}\langle\psi(t)|\hat C^\dagger_{\mu}\hat C_{\mu}|\psi(t)\rangle. $
If $u>\dd p$, no quantum jump occurs, and the unnormalized state $|\widetilde\psi(t)\rangle$ evolves as
\begin{align}
i\frac{\dd}{\dd t}|\widetilde\psi(t)\rangle
&=\Heff|\widetilde\psi(t)\rangle,
\nonumber\\
\Heff&=\hat H-\frac{i}{2}\sum_\mu \hat C_\mu^\dagger \hat C_\mu,
\label{eq:heff}
\end{align}
where $\Heff$ is the non-Hermitian effective Hamiltonian, which describes the simultaneous Hermitian evolution of the system according to $H$ and the non-Hermitian no-jump backaction. The state is then normalized, $\ket{\psi(t+\dd t}=\ket{\tilde\psi(t+\dd t}/\lVert\ket{\tilde\psi(t+\dd t}\rVert$.
Otherwise, if $u\leq\dd p$, a quantum jump is recorded, and the wavefunction is updated as
\begin{equation} \label{eq:quantum-jump}
	|\psi(t + dt)\rangle =  \hat C_\mu 	|\widetilde\psi(t )\rangle /  \|\hat C_\mu|\widetilde\psi(t)\rangle\|\,,
\end{equation}
with the index $\mu$ of the jump $\hat C_\mu$ sampled from the conditional distribution \begin{equation} P(\mu|\mathrm{jump}) = \frac{\langle\psi(t)|\hat C_\mu^\dagger \hat C_\mu|\psi(t)\rangle} {\sum_\nu \langle\psi(t)|\hat C_\nu^\dagger \hat C_\nu|\psi(t)\rangle} \end{equation} by drawing a second random number~\cite{dalibard1992,dum1992,daley2014}.

Rather than evaluating the jump probability at each time step, one can introduce the survival probability as the norm of $\ket{\tilde\psi(t)}$~\cite{dalibard1992,dum1992} 
\begin{equation} \label{Eq:definition_survival}
    s(t) = \| \widetilde\psi(t)\|^2 = \expval{e^{i\Heff^\dagger t}e^{-i\Heff t}}{\psi(0)}\,.
\end{equation}
A single random number $u\in[0,1)$ is extracted, and then the system is evolved according to Eq.~\eqref{eq:heff} until the crossing $s(t)<u$, when a quantum jump acts according to Eq.~\eqref{eq:quantum-jump}.
This algorithm speeds up the evolution of Eq.~\eqref{eq:heff} but still requires evaluating $s(t)$.
This implementation of standard counting quantum trajectories (Monte Carlo wavefunctions) is used, for example, in the
\code{mcsolve} routines of
Refs.~\cite{johansson2012,johansson2013,lambert2024,mercurio2025,guilmin2025}. 
For brevity, we will refer to this as the \textit{standard unraveling}.

To measure the rate at which any quantum jump is expected to interrupt the non-Hermitian Hamiltonian dynamics for an arbitrary wavefunction $\psi$, we can introduce the event activity as
\begin{equation}
	A(\psi)=\sum_\mu
	\langle \psi| \hat C_\mu^\dagger \hat C_\mu|\psi\rangle \equiv \sum_\mu\langle \hat C_\mu^\dagger \hat C_\mu \rangle_{\psi}.
	\label{eq:physical_rate}
\end{equation}

\subsubsection{Cost of quantum trajectories}
\label{sec:trajectory_cost}

A standard unraveling implementation incurs three main costs.
\begin{itemize}
    \item The no-jump evolution of
Eq.~\eqref{eq:heff} is integrated with a (possibly adaptive) ODE solver:
each step applies $\Heff$ to the state at cost $O(N^2)$ for a dense matrix\textsuperscript{\ref{fn:sparse}}
of dimension $N$, and the necessary number of steps is set by the fastest timescale of $\Heff$;
\item  each jump time must be located: resolving the crossing of threshold $u$ by the survival probability $s(t)$ adds a few $O(N^2)$ evaluations per quantum jump by a root solve operation\textsuperscript{\ref{fn:sparse}}~\cite{plenio1998,daley2014,radaelli2024};
\item each jump is applied through
Eq.~\eqref{eq:quantum-jump} at a dominant cost of $O(N^2)$ due to matrix--vector products\textsuperscript{\ref{fn:sparse}}, with an additional delay caused by restarting the integrator from the post-jump
state.
\end{itemize}
The first cost grows with the number of steps over the evolution
time, the other two with the event
activity of Eq.~\eqref{eq:physical_rate} integrated over time.

\subsection{Computational challenges of metastability: the Kerr resonator}
\label{sec:kerr_example}

To highlight the issues at hand, let us consider the coherently driven and dissipative Kerr resonator~\cite{Drummond1980}.  
Its Hamiltonian and dissipator, in the frame rotating at the drive frequency, are
\begin{align}
  \hat H&=-\Delta \hat a^\dagger \hat a+\frac{K}{2}\hat a^{\dagger2}\hat a^2
  +i\left(\varepsilon \hat a^\dagger-\varepsilon^*\hat a\right),
  \nonumber\\
  \hat C&=\sqrt\kappa\,\hat a .
  \label{eq:kerr}
\end{align}
Here, $\hat a$ ($\hat a^\dagger$) is the bosonic annihilation (creation) operator, $\Delta$ is the drive-to-cavity detuning, $K$ is the Kerr nonlinearity, $|\varepsilon|$ is the drive amplitude, and $\kappa$ is the single-photon loss rate.

\begin{figure}[!t]
    \centering
    \reportfig{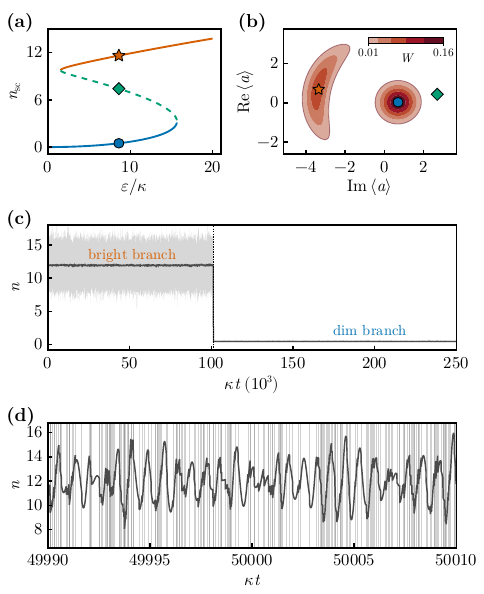}{\columnwidth}
    \caption{Characterization of the driven--dissipative Kerr resonator introduced in Sec.~\ref{sec:kerr_example}.
    \textbf{(a)} Number of photons according to the semiclassical solution in Eq.~\eqref{eq:kerr_mean_field} as a function of the drive $\varepsilon$, with the stable dim (blue) and bright (orange) branches and the unstable saddle
    (dashed green).  
    The bistable window is
    $1.563 \leq \varepsilon/\kappa \leq 15.670$.
    The markers indicate the working point used in the next panels.
    \textbf{(b)} Wigner function of the steady state of Eq.~\eqref{eq:kerr}.
    The orange, blue and green markers indicate the position of the bright, dim, and saddle solutions of Eq.~\eqref{eq:kerr_mean_field}, respectively.    
    \textbf{(c)} Photon number $n(t) = \expval{\hat a^\dagger \hat a}{\psi(t)}$ (light gray) as a function of time for a single trajectory initialized in the bright state predicted by the semiclassical equation of motion. 
    The black solid line is a running average of this same trajectory over time bins of size $100/\kappa$ and serves as a guide to the eye for identifying the switch to the dim phase.
    \textbf{(d)} Zoom on a time window of panel (c) where the system is in the bright branch.
    Vertical lines indicate the times at which quantum jumps occur.
    Parameters: $\Delta/\kappa=13$, $K/\kappa=1.33$, and in panels (b-d)
    $\varepsilon/\kappa=8.616$.
    }
    \label{fig:kerr_bistable}
\end{figure}

This system shows bistability, as can be readily understood using the semiclassical approach (i.e., a coherent-state approximation).
Assume that the density matrix is $\hat\rho = \ketbra{\alpha_{\rm sc}}$, with $\ket{\alpha_{\rm sc}}$ the coherent state such that $\hat a \ket{\alpha_{\rm sc}} = \alpha_{\rm sc} \ket{\alpha_{\rm sc}}$.
At the steady state, where $\dot{\hat\rho} =0$, the equation for $\alpha_{\rm sc} = \expval{\hat a}{\alpha_{\rm sc}}$ reads
\begin{equation}
  \left[\frac{\kappa}{2}-i(\Delta-Kn_{\rm sc})\right]\alpha_{\rm sc}=\varepsilon,
  \qquad n_{\rm sc}=|\alpha_{\rm sc}|^2.
  \label{eq:kerr_mean_field}
\end{equation}
We plot the photon number $n_{\rm sc}$ according to the semiclassical solution as a function of $\varepsilon$ in  Fig.~\ref{fig:kerr_bistable}(a).
At small drive the system is close to vacuum and $n_{\rm sc} \simeq \varepsilon^2/(\Delta^2 + \kappa^2/4)$.
This is the dim (low-occupation) branch.
At large drive the system is dominated by Kerr nonlinearity and $n_{\rm sc} \simeq (\varepsilon/K)^{2/3}$ in the bright (high-occupation) branch.
In the intermediate (bistable) regime, both bright and dim branches coexist. 
Going beyond the semiclassical prediction and reintroducing the effect of quantum fluctuations transforms these stable states into metastable ones, which ultimately converge towards the unique quantum steady state~\cite{Drummond1980,bartolo2016} [shown in Fig.~\ref{fig:kerr_bistable}(b)] but on very long timescales~\cite{Casteels2017}.

We now investigate what happens to a quantum trajectory initialized in the bright branch in the bistability regime.
In Fig.~\ref{fig:kerr_bistable}(c) we plot the photon number $n = \expval{\hat a^\dagger \hat a}{\psi(t)}$ as a function of time.
First, the system remains in the high-occupation region for a time much longer than $\kappa^{-1}$ before a switch to the dim branch takes place\footnote{The long switching times are associated with the slowest
decaying Liouvillian mode, whose decay rate defines the Liouvillian
gap~\cite{macieszczak2016,Casteels2017}.}.
Second, the rate of quantum jumps is very high, as can be seen in Fig.~\ref{fig:kerr_bistable}(d).
Indeed, Eq.~\eqref{eq:physical_rate} gives $A(\psi) \simeq \kappa\bar n$, with $\bar n$ the average photon number in the bright branch. 
Each quantum jump then carries little information about the presence or absence of a switch to the low branch.
Furthermore, each quantum jump perturbs the state, as is visible from the large, persistent, and irregular oscillations in the photon number.

All three costs identified in Sec.~\ref{sec:trajectory_cost} are unfavorable in this regime. 
The no-jump integration must resolve the oscillations of $\Heff$, so the ODE stepper is forced to take many small steps within each photon lifetime $\kappa^{-1}$. 
This inefficiency is compounded by the large number of jumps.
Finally, the long switching time requires large values of $T$ to observe even a handful of switches and properly characterize this process. 

\section{The \mbox{DiSLOU} construction}
\label{sec:method}

In the following section, we detail the working principles of \mbox{DiSLOU}.

\subsection{Layer~I: locally optimal unraveling}
\label{sec:branch}

Layer I aims to minimize the number of quantum jumps occurring during the evolution by finding the optimal unraveling for the states in close phase-space proximity to each metastable attractor.

\subsubsection{Unravelings and gauges}
\label{sec:unraveling}

In the quantum trajectories picture, there is a gauge freedom for the non-Hermitian Hamiltonian describing the time evolution of the system~\cite{wiseman2009,breslin1995,lucas2013,lucas2014}.
For an arbitrary \textit{gauge}, i.e., a set of complex scalars $g = \{ \zeta_\mu^{(g)} \}$, the transformation
\begin{align}
 \hat C_\mu ^{(g)} &= \hat C_\mu+  \zeta_\mu^{(g)} \mathbbm{1},
  \nonumber\\
  \hat H^{(g)}&=\hat H+\frac{i}{2}\sum_\mu
  \left(\zeta_\mu^{(g)} \hat C_\mu^\dagger-\zeta_\mu^{(g)\,*}\,\hat C_\mu\right)\,,
  \label{eq:scalar_shift}
\end{align}
where $\mathbbm{1}$ is the identity matrix in the $N$-dimensional Hilbert space, leaves the Lindblad master equation of Eq.~\eqref{eq:lindblad} unchanged, but it defines a new unraveling.  
The resulting effective Hamiltonian in Eq.~\eqref{eq:heff} becomes
\begin{equation}
  \Heff^{(g)}=\Heff-i\sum_\mu\zeta_\mu^{(g)*}\hat C_\mu
  -\frac{i}{2}\sum_\mu|\zeta_\mu^{(g)}|^2\mathbbm{1}.
  \label{eq:branch_heff}
\end{equation}
No approximation or bias is introduced by changing the gauge, and averaging over many quantum trajectories leads to the same results, independently of the gauge.
However, the dynamics of single quantum trajectories change. For example, the activity transforms as
\begin{align}
	A^{(g)}(\psi)
	& = \sum_\mu \langle \hat C_\mu ^{(g) \,\dagger} \hat C_\mu ^{(g)} \rangle_{\psi}=
	\nonumber\\
    &=\sum_\mu\left[
	\Var_\psi(\hat C_\mu^{(g)})
	+|\langle \hat C_\mu\rangle_\psi+\zeta_{\mu}^{(g)}|^2
	\right],
\label{eq:gauge_rate}
\end{align}
where
$\Var_\psi(\hat C)=\langle \hat C^\dagger \hat C\rangle_\psi
-|\langle \hat C\rangle_\psi|^2$ is the variance of the operator $\hat C$ for a wavefunction $\ket{\psi}$.

While the gauge transformation is an algebraic identity for arbitrary channels, in the case of single-photon loss in, e.g., Eq.~\eqref{eq:kerr}, it also has a direct experimental interpretation. 
Since photon loss can be interpreted as the monitoring of the output field, changing the gauge implies the interference of the output field with a coherent local oscillator of amplitude $\zeta_\mu^{(g)}$ before detection, as sketched in Fig.~\ref{fig:monitoring_geometry}(a).
Such a protocol, for large $\zeta_\mu^{(g)}$, is known as homodyne detection~\cite{milburn1993}.

\subsubsection{Optimal unraveling and finite gauge set}

If we assume the variance of the jump operators to be gauge-independent, the activity $A^{(g)}(\psi)$ of Eq.~\eqref{eq:gauge_rate} can be minimized along a trajectory by choosing the instantaneous gauge $\zeta_\mu^{(g)}(t)=-\langle \hat C_\mu\rangle_{\psi(t)}$.
While possible in principle, this protocol incurs an additional $O(N^2)$ cost to evaluate $\langle \hat C_\mu\rangle_{\psi(t)}$ at each time step.
Such a construction would also prevent the use of the diagonal evolution discussed in Layer~II below.

We resolve this with a pragmatic approach: rather than seeking an instantaneous optimum, we restrict the shifts to a finite set of $N_g$ gauges, $\left\{{\zeta_\mu^{(g)}}\right\}_{g=1}^{N_g}$, fixed in advance. 
In practice, we choose roughly one gauge per metastable region of phase-space, with shifts selected to approximately cancel the corresponding jump amplitudes. 
The gauges can be chosen using any numerically efficient criterion\footnote{Note that suboptimal choices of the gauges will raise the number of quantum jumps, and consequently the numerical cost of simulations, but they will not affect correctness.}: an option is to use the semiclassical basins of attraction.
Another possibility (see App.~\ref{app:gauge-searching}) is to perform short time evolutions of a few quantum trajectories without resorting to the optimal unraveling, and wait for them to converge to the metastable regions. 
Then a clustering algorithm can be applied to determine the optimal shifts.

It remains to decide which gauge is used.
At the beginning of a trajectory, we score the candidates by their activity and we choose the gauge $g_0$ for which the activity $A_{g_0}(\psi(0))$ in Eq.~\eqref{eq:gauge_rate} is minimum. 
That is
\begin{equation}
    g_0=\operatorname*{argmin}_g A_g(\psi(0)).
\end{equation}
Once a gauge is fixed, we remain in that gauge until a quantum jump occurs.
Immediately after a jump, the state is perturbed, and we change from a gauge $g$ to
$g'$ if
\begin{equation}
  A_{g'}(\psi(t))<\eta A_{g}(\psi(t)),
  \qquad 0<\eta\le1.
  \label{eq:hysteresis}
\end{equation}
The hysteresis factor $\eta$ prevents repeated gauge switching near a boundary between two attraction basins.
Notice that the state remains unchanged when the selected unraveling changes: only its representation varies.

Figure~\ref{fig:monitoring_geometry}(b,c) investigates the dynamics of the bright branch of the Kerr resonator in the optimal gauge minimizing the activity in Eq. \eqref{eq:gauge_rate}.
For quantum jumps to be less probable, the survival probability $s(t)$ in Eq.~\eqref{Eq:definition_survival} must decay at a lower rate.
This is shown in Fig.~\ref{fig:monitoring_geometry}(b), where we compare $s(t)$ in the original gauge and in the optimal one.
Starting from the same initial coherent state obtained from the semiclassical solution in Eq.~\eqref{eq:kerr_mean_field}, it now takes around one order of magnitude longer for $s(t)$ to cross the survival threshold.
We then plot in Fig.~\ref{fig:monitoring_geometry}(c) the photon number dynamics over the same time interval as in Fig. \ref{fig:kerr_bistable}(d).
The measurement record in the original gauge carries $223$ jump counts, while this number drops to $35$ in the locally optimal gauge.
Not only are quantum jumps fewer, but they tend to bunch together, resulting in long timespans of fairly regular non-Hermitian dynamics. 
We will make use of this property for Layer III.

\subsection{Layer II: diagonal propagation and first passage}
\label{sec:full_spectral}

Even after having minimized the number of quantum jumps, the switching events remain rare, thus requiring very long evolutions to sample them.
Therefore, it makes sense to reduce the time spent evolving the state between two quantum jumps.
Layer II addresses this problem.

\subsubsection{Construction of the diagonal evolution}

To explain the idea behind the diagonal evolution approach, let us assume we are working with a single gauge, so that we can drop the $g$ label.
The no-jump generator in Eq.~\eqref{eq:heff} is time independent, so
the unnormalized wavefunction evolves as 
\begin{equation}
    |\widetilde\psi(t)\rangle=e^{-i\Heff
t}|\widetilde\psi(0)\rangle.
\label{eq:no_jump_evolution}
\end{equation}
Let the effective Hamiltonian be diagonalizable and
\begin{equation}
\begin{split}
  \Heff V=V\Lambda,
   & \qquad
  \Lambda=\diag(\lambda_1,\ldots,\lambda_N), \\
  &V = \{v_1,\ldots,v_N\},
  \end{split}
  \label{eq:eigen}
\end{equation}
where the columns $v_j$ of $V$ are normalized right eigenvectors.
In this eigenbasis, the no-jump propagation is diagonal and
\begin{equation}
  c(t)=D(t)c(0),
  \qquad
  D(t)=\diag\!\left(e^{-i\lambda_jt}\right),
  \label{eq:diagonal_flow}
\end{equation}
where $c(t)$ are the coefficients representing $|\widetilde\psi(t)\rangle$ in the basis of the eigenvectors.
Thus, after an initial diagonalization of the non-Hermitian Hamiltonian at cost $\mathcal{O}(N^3)$, advancing the coordinates to any later time costs only $\mathcal{O}(N)$ independently of $t$\footnote{The use of the diagonal basis $\Heff$ in \code{QuantumToolbox.jl} was \href{https://github.com/qutip/QuantumToolbox.jl/issues/341}{originally introduced} by Adrià Labay-Mora, and used for the simulations in Ref.~\cite{labaymora2026chiralcat}.}.

There are, however, penalties in moving to the diagonal basis. 
$\Heff$ is non-Hermitian and, in general, nonnormal, so that the eigenvectors $v_j$ are not mutually orthogonal and $V$ is a non-unitary dense matrix.
Observables, jump operators, and the norm of the wavefunction are defined in the original basis, where they are often sparse.
We therefore reconstruct the wavefunction $|\widetilde\psi(t)\rangle=Vc(t)$ whenever it is required, through a dense matrix--vector product of cost $O(N^2)$, and evaluate these quantities with the original operators.
Conversely, after each quantum jump, the coordinates $c=V^{-1}|\psi\rangle$ of the new state must be computed.
Rather than inverting $V$, we use its QR decomposition $V=QR$, with $Q$ unitary and $R$ upper triangular, so that $c=R^{-1}Q^\dagger|\psi\rangle$ requires one matrix--vector product and one triangular solve, again at cost $O(N^2)$.
In terms of the coordinates, the inner product between wavefunctions reads
\begin{equation}
  \langle\psi|\psi\rangle=c^\dagger Gc,
\qquad
  G=V^\dagger V=R^\dagger R\,,
  \label{eq:metric}
\end{equation}
where the off-diagonal entries of the Gram matrix $G$ carry the overlaps between eigenmodes.
Therefore, the survival probability in Eq.~\eqref{Eq:definition_survival} reads
\begin{equation}
  s(t)=\lVert VD(t)c\rVert^2=c^\dagger D^\dagger(t)GD(t)c.
  \label{eq:survival}
\end{equation}
Evaluating Eq.~\eqref{eq:survival} at any requested time costs $O(N^2)$ operations, as $V$ is dense.

These trade-offs are more than counterbalanced by the fact that propagating the wavefunction and then computing $s(t)$ costs only $O(N^2)$ overall instead of the original $O(N^2)$ per time step: for the diagonal evolution propagating over a long no-jump interval is no more expensive than over a short one.
Thus, the waiting-time construction of Sec.~\ref{sec:unraveling_geometry} can be carried out without any step-by-step evolution, but rather by directly finding the first passage time of the survival probability at the random threshold $u\in[0,1)$
\begin{equation}
  \tau=\inf\{t\ge0:s(t)\le u\}\,.
  \label{eq:first_passage}
\end{equation}
This operation can be efficiently performed in an interval $[0,T]$ by a bracketing root solve, where every iteration involves an independent $O(N^2)$ evaluation of Eq.~\eqref{eq:survival}.
At the crossing, the appropriate jump operator is drawn by a second random number extraction, as done in Sec.~\ref{sec:standard_unraveling} for the standard unraveling.
Appendix~\ref{app:root-solve} describes in more detail how the jump time is located in our implementation.

Although the method works, in principle, for any diagonalizable $\Heff$, models with strongly ill-conditioned $V$ would require a different treatment (e.g., a Schur decomposition rather than a diagonalization).
Indeed, when eigenvectors become nearly linearly dependent, the coordinates $c=R^{-1}Q^\dagger|\psi\rangle$ lose up to $\log_{10}\kappa(V)$ significant digits, with $\kappa(V)$ the condition number of $V$, although physical states typically lose far fewer.
Our implementation therefore warns when $\kappa(V)\geq\epsilon_{\rm mach}^{-1/2}$, with $\epsilon_{\rm mach}$ the machine precision, i.e., when up to half of the significant digits can be lost, and stops when $\kappa(V)\geq\epsilon_{\rm mach}^{-1}$, i.e., when $\Heff$ is numerically not diagonalizable, as at an exceptional point.

\begin{figure}[t]
  \centering
  \reportfig{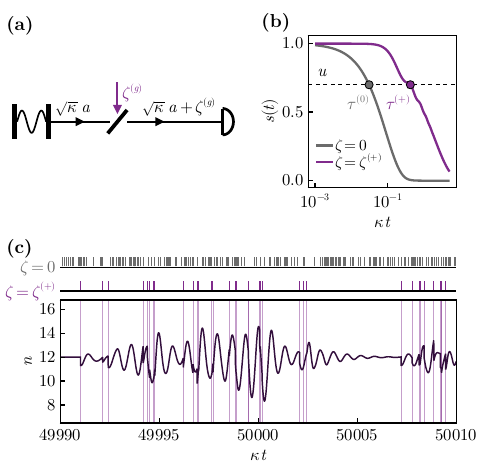}{\linewidth}
  \caption{
  Dynamics of the Kerr resonator in the bright branch at the same bistable point as in  Fig.~\ref{fig:kerr_bistable}, but in the gauge $\zeta =\zeta^{(+)}$ that minimizes the activity [see Eq.~\eqref{eq:gauge_rate}].
\textbf{(a)}~Sketch of a homodyne detection apparatus, where the cavity output field (associated with the single-photon loss jump operator $\sqrt{\kappa}\hat{a}$) is mixed by a beamsplitter with a coherent local oscillator of amplitude $\zeta^{(g)}$ prior to detection.
Monitoring the resulting output port corresponds to the shifted jump operator $\sqrt\kappa\hat{a}+\zeta^{(g)}\mathbbm{1}$, i.e., the transformation described in Eq.~\eqref{eq:scalar_shift}.
  \textbf{(b)}~Survival probability in Eq.~\eqref{eq:survival} for an initial coherent state with the amplitude predicted by the semiclassical solution in Eq.~\eqref{eq:kerr_mean_field}.
  The dashed line indicates the value of one draw $u$ and the circles mark the first passage time $\tau^{(g)}$.
  In the chosen gauge, the first passage is $13.7$ times longer than in the original $\zeta=0$ gauge.
  \textbf{(c)}~As in Fig.~\ref{fig:kerr_bistable}(d), photon number along a trajectory, with the vertical purple lines marking the quantum jumps. 
  Notice that in this case the jump operator is $\hat{C}^{(g)} = \hat{a} +  \zeta^{(+)}\mathbbm{1}$.
  Above the panel we illustrate the sequence of the quantum jumps in the original ($\zeta=0$) and optimal ($\zeta = \zeta^{(+)}$) gauges.
  Parameters as in Fig.~\ref{fig:kerr_bistable}.
  }
\label{fig:monitoring_geometry}
\end{figure}

\subsubsection{Integrating Layer~I and II}

Notice that, despite the advantages of using a first-passage formalism, each jump event still requires a root solve at $O(N^2)$ per iteration
and one application of the respective operator, as in Eq.~\eqref{eq:quantum-jump}.
Both scale with the number of quantum jumps, which motivates combining this method with Layer~I. The whole construction in Sec.~\ref{sec:full_spectral} applies verbatim to each $\Heff^{(g)}$, and their spectral representation is prepared once and reused for every trajectory and every interval assigned to $g$.
The resulting algorithm has a set of $N_g$ effective diagonalized Hamiltonians,  replacing the cost of evolution between quantum jumps
detailed in Sec.~\ref{sec:trajectory_cost} with a small number $N_g$ of $O(N^3)$ diagonalizations, whose expense is amortized over every evolution between two quantum jumps of every quantum trajectory.

\subsection{Layer~III: local reduced eigenspaces}
\label{sec:reduced}

A state that remains close to a metastable attractor and that maintains a small variance in the activity populates only a handful of the eigenmodes of the corresponding gauge. 
Layer III makes use of this property to further speed up the calculations.

\begin{figure}[!t]
  \centering
  \reportfig{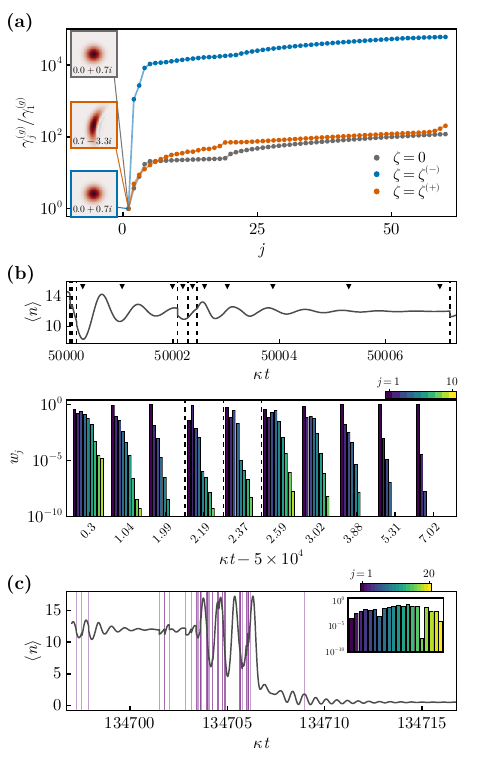}{\linewidth}
  \caption{
  Properties of the spectrum and eigenvectors of the non-Hermitian Hamiltonians $H_{\rm eff}^{(g)}$ in different gauges.
  \textbf{(a)}~Eigenvalue decay rates $\gamma_{j}^{(g)}$ for the undisplaced
  gauge ($\zeta=0$) and for the two shifted ones ($\zeta=\zeta^{(\pm)}$)
  , each
  normalized to its own smallest activity $\gamma_1^{(g)}$. 
  Insets show the Wigner functions of the least-decaying eigenmode of
  each $\Heff^{(g)}$ together with their amplitudes $\expval{\hat{a}}$ which roughly corresponds to $-\zeta^{(g)}$. Note that these Wigner functions resemble the metastable states identifiable from Fig.~\ref{fig:kerr_bistable}(b).
  \textbf{(b)} Top: zoom on the dynamics in Fig.~\eqref{fig:monitoring_geometry}, i.e., a system in the bright branch evolved using the optimal gauge $\zeta = \zeta^{(+)}$, which minimizes the activity.  
  The expectation value of the photon number $n$ with respect to time is plotted, with the quantum jumps indicated by dashed lines.
  Triangle markers indicate the ten time points used in the bottom panel.
  Bottom: histograms of the weights $w_j=|c_j|^2/\sum_k|c_k|^2$, where $c_j$ is the $j$-th coefficient of the coordinate vectors of Eq.~\eqref{eq:diagonal_flow}, of the ten slowest decaying (smallest $\gamma_{j}^{(g)}$) modes. 
  Vertical lines indicate that a jump occurred between two consecutive histograms.
  \textbf{(c)}~One trajectory evolved using Layer~I around a branch switch: the photon number (dark line) with its quantum jump events (vertical lines) in the optimized gauge.  The inset gives the weights of the $20$ slowest (smallest $\gamma_j$) modes at time $\kappa t\simeq 134706$, immediately after the last jump before the drop in the photon number, revealing a significant spread of the occupation across all states shown.
  Parameters as in Fig.~\ref{fig:kerr_bistable}.}
  \label{fig:conditional_spectrum}
\end{figure}

\subsubsection{Relevant eigenvalues between two quantum jumps}

Let us consider a system in the gauge $g$ that minimizes its activity.
How does an eigenstate $v_j^{(g)}$ of the non-Hermitian Hamiltonian evolve between two quantum jumps?
Using $\Heff^{(g)} v_j^{(g)}=\lambda_j^{(g)}v_j^{(g)}$ and the evolution rule of Eq.~\eqref{eq:heff}, we have $\frac{\dd}{\dd t}\lVert v_j^{(g)}\rVert^2=2\Im\lambda_j^{(g)}\equiv-\gamma_j^{(g)}$.
We call $\gamma_j^{(g)}\geq0$ the eigenvalue decay rate for a given gauge $g$\footnote{Note that this also corresponds to the activity of Eq.~\eqref{eq:physical_rate} evaluated on eigenmode $j$: $A(v_j) =\sum_\mu\|\hat{C}_\mu v_j\|^2  = -2\,\operatorname{Im}\lambda_j=\gamma_j$.}: it is the rate at which the norm of the corresponding eigenmode is lost between jumps. 
As jumps are now rare and consequently no-jump intervals long, fast decaying modes will contribute little to the dynamics.

We confirm this intuition about the role of the rapidly decaying eigenmodes in Fig. \ref{fig:conditional_spectrum}.
In particular, in Fig. \ref{fig:conditional_spectrum}(a) we plot the eigenvalue decay rates of the bistable Kerr resonator model for the unmodified $\Heff$ ($\zeta = 0$) and for the optimal gauges corresponding to the dim and bright branches ($\zeta = \zeta^{(-)}$ and $\zeta = \zeta^{(+)}$, respectively). 
It is evident that, for the optimized gauges, most eigenmodes are quickly decaying, and they separate from a much smaller number of low activity modes. 
The latter can survive long evolutions under the effective Hamiltonian without undergoing a jump.
In Fig.~\ref{fig:conditional_spectrum}(b), we show a section of the quantum trajectory already shown in Fig.~\ref{fig:monitoring_geometry}(c), while also marking the times for which we plot the weight of each eigenvector of the non-Hermitian Hamiltonian in the gauge $\zeta = \zeta^{(+)}$.
During propagation under $\Heff^{(g)}$, only the very first modes maintain sizable weights, while after a quantum jump the higher modes get repopulated.

\subsubsection{Construction of the reduced basis}

Following these observations, Layer~III performs, when possible, a truncation of the local Hilbert space around the $m^{(g)}\ll N$ least-decaying modes of $\Heff^{(g)}$, and evolves the wavefunction of these few modes. 
Evolution in the reduced basis has leading cost $O(m^{(g)\:2})$ rather than $O(N^2)$.

To select the eigenvectors of $\Heff^{(g)}$ to retain, we order them by increasing eigenvalue decay $\gamma_j^{(g)}$.
We then keep the first $m^{(g)}$, either by fixing their number or by specifying a cutoff.
The basis reads
\begin{equation}
  \mathcal K^{(g)}_{m}=\operatorname{span}
  \{v^{(g)}_{1},\ldots,v^{(g)}_{m}\}.
  \label{eq:slow_subspace}
\end{equation}
This operation has no additional cost as Layer~II already produced the eigendecomposition of $\Heff^{(g)}$.

Propagating inside $\mathcal K_m^{(g)}$ is achieved by applying the machinery of Layer~II to the $m^{(g)}$ retained state coordinates. Since $\mathcal K_m^{(g)}$ is invariant under $\Heff^{(g)}$, the diagonal propagator is obtained by restricting its full-space counterpart to the retained modes, $D_m^{(g)}(t)$.
To evaluate norms and observables without reconstructing the $N$-dimensional wavefunction, we introduce the thin QR decomposition of the $N\times m^{(g)}$ matrix $V_m^{(g)}$ of the retained eigenvectors,
\begin{equation}
  V_m^{(g)}=Q_m^{(g)}R_m^{(g)},
  \qquad
  Q_m^{(g)\dagger}Q_m^{(g)}=\mathbbm{1},
  \label{eq:reduced-qr}
\end{equation}
with $R_m^{(g)}$ an $m^{(g)}\times m^{(g)}$ upper-triangular matrix.
A state $|\psi\rangle=V_m^{(g)}c$ is then stored through its orthonormal coordinates $q=Q_m^{(g)\dagger}|\psi\rangle=R_m^{(g)}c$, whose Euclidean norm equals $\lVert\psi\rVert$, so that the survival probability requires no Gram matrix.
Observables are represented by the $m^{(g)}\times m^{(g)}$ matrices $O_m^{(g)}=Q_m^{(g)\dagger}\hat{O}Q_m^{(g)}$, computed once.

This is a low-rank approximation of the dynamics around each metastable attractor. Notice that it bears a resemblance to the local dynamics of the density matrix, as the selected modes are often similar to, but can differ from, the slow eigenvectors of the Liouvillian used to describe metastable manifolds~\cite{macieszczak2016,minganti2018}.

\subsubsection{Gauge change and fallback criterion}

Within Layer~III, we spend most of the time in the reduced basis. 
However, complex dynamics such as the switching event between metastable attractors often require the full basis to be correctly described.
This is shown in Fig. \ref{fig:conditional_spectrum}(c), where the event activity increases dramatically close to the switching event, redistributing weight over a large number of eigenmodes (inset).
Therefore, we must be capable of cheaply identifying when it is safe to
work in the reduced basis and when we need to fall back to the full one.

First, we remark that a state that starts inside the space $\mathcal K^{(g)}_{m}$ cannot leak out
of it by Eq.~\eqref{eq:diagonal_flow} because $\mathcal K^{(g)}_{m}$ is spanned by eigenvectors of $\Heff$. 
It is only after a jump, where neither the population of the discarded states increases nor a gauge change takes place, that we need to assess if the full basis is required. 
As these are also the times when we verify the optimal gauge for Layer~I, this procedure does not further interrupt the diagonal evolution of Layer~II.

To identify the fallback condition, let us introduce the projector $P^{(g)}_{m}=V_m^{(g)}\left[V_m^{(g)\dagger}V_m^{(g)}\right]^{-1}V_m^{(g)\dagger}$ onto the subspace $\mathcal K^{(g)}_{m}$.
Given a normalized wavefunction $\ket{\psi}$ and its projection $\ket{\psi_m^{(g)}}=P_m^{(g)}\ket{\psi}/\|P_m^{(g)}\ket{\psi}\|$, we define the residuals
\begin{equation}
  r^{(g)}_{m}(\psi)=
  \|(I-P^{(g)}_{m})\psi\| = D_{\rm tr}\!\left(\ket{\psi}, \ket{\psi_m^{(g)}}\right),
  \label{eq:residual}
\end{equation}
with the trace distance~\cite{nielsen2010quantum} defined as $D_{\rm tr}(\ket{\psi},\ket{\phi})=\sqrt{1-\lvert\braket{\psi}{\phi}\rvert^2}$.
We fall back to the full basis whenever $r^{(g)}_{m}(\psi) > r_{\rm tol}$, with the tolerance $r_{\rm tol}$ fixed in advance. The typical values of $r_{\rm tol}$ we employ are between $10^{-3}$ and $10^{-4}$.
Equation~\eqref{eq:residual} bounds the maximum error introduced on observables at one
projection.
It does not, however, bound the error over the whole trajectory as later normalization, jump selection, and nonnormal propagation can amplify errors. 
Therefore, careful verification of convergence in $r_{\rm tol}$ for each problem is essential.

There are changes of gauges, say from $g \to g'$, that do not require passing through the full basis. 
If the dynamics lies at first within the reduced basis $\mathcal K_m^{(g)}$ and then moves to that of $\mathcal K_{m'}^{(g')}$, the combined action of the jump operator and the change of basis associated with the change of gauge acts on the coefficients of the wavefunction as $c' = J_{\mu,\mathrm{red}}^{g'\leftarrow g} c$ with
\begin{equation}
J_{\mu,\mathrm{red}}^{g'\leftarrow g}
=
\left[
V_{m'}^{(g')\dagger}V_{m'}^{(g')}
\right]^{-1}
V_{m'}^{(g')\dagger}
\hat{C}_\mu^{(g)}
V_m^{(g)}.
\label{eq:reduced-transfer}
\end{equation}
In this case, the $N$-dimensional full-basis vector is never formed.
Appendix~\ref{app:reduced-jump} provides more details on this procedure.

\subsection{Solver workflow}
\label{sec:hierarchy}

% Schematic of the DiSLOU solver loop, Sec.~\ref{sec:hierarchy}.
% Requires tikz with libraries arrows.meta, positioning, calc, shapes.geometric.
\definecolor{layI}{HTML}{2C6FAF}
\definecolor{layII}{HTML}{0E8074}
\definecolor{layIII}{HTML}{BC7412}
\begin{figure}[!t]
\centering
\begin{tikzpicture}[
  font=\scriptsize,
  >={Latex[length=1.3mm,width=1.1mm]},
  box/.style={draw=#1,fill=#1!7,line width=0.45pt,rounded corners=1.8pt,
              minimum width=3.9cm,minimum height=5.2mm,inner xsep=3pt,align=center},
  test/.style={diamond,aspect=3.1,draw=#1,fill=#1!14,line width=0.8pt,
               inner xsep=1pt,inner ysep=0pt,minimum height=5.2mm},
  pill/.style={draw=#1!60,fill=white,line width=0.45pt,rounded corners=3pt,
               minimum width=1.4cm,minimum height=4.6mm,inner xsep=2pt,align=center},
  bdg/.style={draw=#1,fill=#1!12,text=#1,line width=0.4pt,rounded corners=1.2pt,
              font=\tiny\bfseries,inner xsep=2pt,inner ysep=1.2pt,minimum width=5.6mm},
  ar/.style={->,draw=black!60,line width=0.45pt},
  ln/.style={draw=black!60,line width=0.45pt},
  tag/.style={font=\tiny,text=black!55,inner sep=1.2pt},
  cage/.style={draw=black!25,dash pattern=on 2.6pt off 2pt,rounded corners=3pt,
               line width=0.4pt,inner xsep=4pt,inner ysep=6pt},
  ttl/.style={font=\tiny,text=black!50,fill=white,inner xsep=2.5pt,inner ysep=0.8pt}
]
\newcommand{\hd}[1]{{\tiny\textcolor{black!55}{#1}}\\[0.2ex]}
\begin{scope}[
]
\def\rail{-3.95}
\def\side{2.65}
\def\loop{3.75}
\def\vs{4.2mm}

\node[box=layII]  (spec) at (0,0) {\hd{find gauges, calculate spectra}$\{\zeta^{(g)}\}_{g=1}^{N_g}$, $\{\lambda^{(g)}_j,v^{(g)}_j\}$};
\node[box=layI,below=\vs of spec]  (sel) {\hd{pick the gauge}$g_\star=\operatorname*{argmin}_g A_g$};
\node[test=layIII,below=\vs of sel] (t1) {$r\le r_{\rm tol}$};
\node[pill=layIII] (k1) at (-\side,0|-t1) {\hd{reduced}$\mathcal K^{(g_\star)}_m$};
\node[pill=layIII] (f1) at ( \side,0|-t1) {\hd{full space}$\mathbb C^N$};

\node[box=layII,below=13mm of t1] (fp) {\hd{root solve}$s(\tau)=u$, $u\in[0,1)$};
\node[box=layII,below=\vs of fp]  (jmp) {\hd{draw the jump channel}$\mu\sim\|\hat C^{(g)}_\mu\psi\|^2$};
\node[box=layI,below=\vs of jmp]  (sw)  {\hd{gauge update}$A_{g'}<\eta A_g\Rightarrow g\to g'$};
\node[test=layIII,below=\vs of sw] (t2) {$r\le r_{\rm tol}$};
\node[pill=layIII] (k2) at (-\side,0|-t2) {\hd{stay reduced}$J^{g'\leftarrow g}_{\mu,\rm red}$};
\node[pill=layIII] (f2) at ( \side,0|-t2) {\hd{fall back}$\hat C^{(g)}_\mu\psi\in\mathbb C^N$};

\node[bdg=layII]  (nspec) at (\rail,0|-spec) {II};
\node[bdg=layI]   (nsel) at (\rail,0|-sel)  {I};
\node[bdg=layIII] (nt1) at (\rail,0|-t1)   {III};
\node[bdg=layII]  (nfp) at (\rail,0|-fp)   {II};
\node[bdg=layII]  (njmp) at (\rail,0|-jmp)  {II};
\node[bdg=layI]   (nsw) at (\rail,0|-sw)   {I};
\node[bdg=layIII] (nt2) at (\rail,0|-t2)   {III};

\draw[ar] (spec) -- (sel);
\draw[ar] (sel) -- (t1);
\draw[ar] (fp) -- (jmp);
\draw[ar] (jmp) -- (sw);
\draw[ar] (sw) -- (t2);

\node[tag,anchor=south west] at ($(t1.north)+(0.3,0.04)$) {residual test};
\draw[ar] (t1.west) -- (k1) node[tag,midway,above] {$\checkmark$};
\draw[ar] (t1.east) -- (f1) node[tag,midway,above] {$\times$};
\coordinate (m1) at ($(t1.south)+(0,-7mm)$);
\draw[ln] (k1.south) |- (m1);  \draw[ln] (f1.south) |- (m1);
\draw[ar] (m1) -- (fp.north);

\node[tag,anchor=south west] at ($(t2.north)+(0.3,0.04)$) {residual test};
\draw[ar] (t2.west) -- (k2) node[tag,midway,above] {$\checkmark$};
\draw[ar] (t2.east) -- (f2) node[tag,midway,above] {$\times$};
\coordinate (m2) at ($(t2.south)+(0,-4.2mm)$);
\draw[ln] (k2.south) |- (m2);  \draw[ln] (f2.south) |- (m2);
\draw[ar] (m2) -| (\loop,0|-jmp) |- (fp.east);
\begin{scope}[on background layer]
  \node[cage,fit=(nspec)(spec)(t1)(k1)(f1)] (cA) {};
  \node[cage,fit=(nfp)(fp)(t2)(k2)(f2)] (cB) {};
\end{scope}
\node[ttl,anchor=west] at ($(cA.north west)+(0.3,0)$) {\textsc{preparation}};
\node[ttl,anchor=west] at ($(cB.north west)+(0.3,0)$) {\textsc{per no-jump interval}};
\end{scope}
\end{tikzpicture}
\caption{Schematic of the \mbox{DiSLOU} solver workflow. Badges on the left indicate the relevant layer.
The steps labeled \emph{preparation} are performed only once per evolution, while the ones labeled \emph{per no-jump interval} are repeated for each no-jump evolution of each trajectory.}
\label{fig:workflow}
\end{figure}
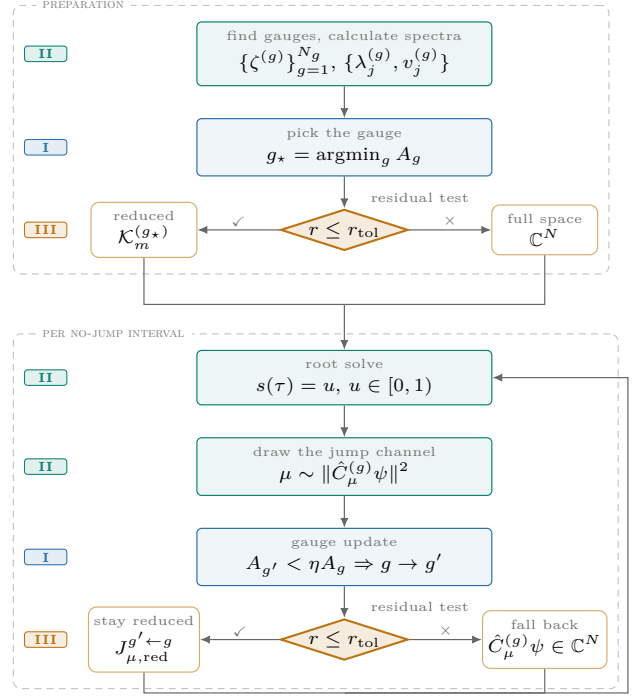

Because the three layers act on different cost factors, their optimal usage depends on the problem at hand.
For instance, 
using Layer~III may actually worsen performance for an evolution that often switches between a reduced basis and a full one. In a different scenario, it may be useful to use Layer~I, but if it fails to reduce the activity enough, for example due to incorrect selection of the gauges, Layer~II can become inefficient.
The realized gain therefore depends jointly on the event activity, the presence of metastable states, the
fraction of time spent in reduced spaces, and the frequency of rejected projections.
Nonetheless, for the cases of interest studied in this article, we \textit{always} find an increase in computational speed, even in non-optimal configurations.

Overall, when all three layers work together, the solver follows the sequence sketched in Fig.~\ref{fig:workflow}.
At initialization, $g_\star=\operatorname*{argmin}_{g'}A^{(g')}$ selects the
gauge, after which Eq.~\eqref{eq:residual} determines whether
propagation begins in $\mathcal K_{g,m_g}$ or in the full space.
During each subsequent no-jump interval the current gauge remains fixed while the first crossing of the threshold $u$ by the survival probability is computed through a single root-solve,
as in Eq.~\eqref{eq:first_passage}.
The sampled shifted channel $\hat{C}_\mu^{(g)}$ produces the normalized post-jump
state.
Only then is $g_\star$ recomputed: the next gauge is $g_\star$ when
$A^{(g_\star)}<\eta A^{(g)}$, and remains $g$ otherwise.
Throughout this process, Layer~III uses the full basis only when necessary, and this is a marginal fraction of the total evolution.
The residual calculated in the destination gauge determines whether the
next interval uses the reduced space or falls back to the full space, for example in the already mentioned case of rare transitions from one attractor to the other.

In the companion implementation of \mbox{DiSLOU}~\cite{DiSLOU_git}, this workflow can be easily invoked, for the case of the driven-dissipative Kerr resonator of Sec.~\ref{sec:kerr_example}, as:
\begin{lstlisting}
using DiSLOUTrajectories
using QuantumToolbox

N = 60
κ, Δ, K = 1.0, 13.0, 1.33
ε = 8.616
ntraj = 512
tlist = range(0.0, 50.0/κ; length=51)

a = destroy(N)
n_op = a' * a
H = -Δ * n_op + (K / 2) * a'^2 * a^2 +
    im * (ε * a' - conj(ε) * a)
c_ops = [sqrt(κ) * a]

sc = discover_gauges(
    H, c_ops;
    method=:semiclassical,
    mode_ops=[a],
    mode_dims=[N],
    limits=(30,),
)

α_saddle = only(sc.diagnostics.rejected)
ψ0 = coherent(N, α_saddle)

layer3_sol = dislou_solve(H, ψ0, tlist, c_ops;
    e_ops=[n_op],
    gauge_set=sc,
    ntraj=ntraj,
    layer3_sizes=[20, 60],
    residual_tolerance=1e-4,
)
\end{lstlisting}
where the gauges are discovered by solving the semiclassical equations and the initial state is set at the position of the only unstable saddle.

The function \code{dislou\_solve} is built on top of the \code{mcsolve} function of \code{QuantumToolbox.jl}~\cite{mercurio2025}.
The no-jump propagation of Layers~II and III is implemented as an exact integrator for the ordinary differential equation solvers of \code{DifferentialEquations.jl}~\cite{rackauckas2017differentialequations}, on which \code{mcsolve} relies.
This integrator also applies the quantum jumps and selects the gauge and the basis of each no-jump interval, while \code{mcsolve} draws the random numbers, locates the jump times (see App.~\ref{app:root-solve}), and runs the ensemble of trajectories.
As a consequence, \code{dislou\_solve} accepts the keyword arguments of \code{mcsolve}, e.g., to set the random number generator, to save the states, or to distribute the trajectories over threads or processes, and returns the same solution object, so that the results are analyzed as for standard quantum trajectories. For instance, \code{layer3\_sol.expect} contains the ensemble average of $\langle\hat n\rangle$ at the times in \code{tlist}.

\section{In-depth benchmark on the driven-dissipative Kerr resonator}
\label{sec:kerr}

We now provide more quantitative data on the performance and accuracy of \mbox{DiSLOU} on the Kerr resonator of Sec.~\ref{sec:kerr_example}. 
For this exercise, we take $\Delta=13\kappa$ and $K=0.2\kappa$, measure time in units of $\kappa^{-1}$, and truncate the Fock space at $N=190$. 
The semiclassical bistable region extends from $\varepsilon_{\rm l}=4.0304\kappa$ to $\varepsilon_{\rm h}=40.408\kappa$: below $\varepsilon_{\rm l}$ only the dim branch is stable, whereas above $\varepsilon_{\rm h}$ only the bright branch remains. 
In the dim and bright monostable regimes we fix the gauge to $\zeta=-\sqrt{\kappa}\alpha$, with $\alpha$ the solution to the semiclassical equation \eqref{eq:kerr_mean_field} at the respective value of $\varepsilon$.
In the bistable regime, instead, we need two gauges, one for the bright and one for the dim attractor. 
Therefore we fix $\zeta^{(\pm)}
=-\sqrt{\kappa}\alpha_{\pm}$, where $\alpha_{\pm}$ are the two stable solutions of Eq.~\eqref{eq:kerr_mean_field}.

We benchmark \mbox{DiSLOU} at three drive amplitudes, corresponding to three different regimes:
\begin{itemize}
    \item the dim monostable regime. We choose $\varepsilon =3.6273 \kappa < \varepsilon_l$. For this set of parameters $\alpha = 0.0107+0.2789i$;
    \item the  bistable regime $\varepsilon_l< \varepsilon =22.217 \kappa < \varepsilon_h$.
    Here, $\alpha_-  = 0.0727+1.7954i$ and $\alpha_+ = 1.7414-8.6224i$.
    \item the bright regime $\varepsilon_h< \varepsilon =44.449 \kappa$, where $\alpha = 0.9954-9.3542i$.
\end{itemize}

\subsection{Accuracy}
\label{sec:kerr_accuracy}

\begin{figure}[!ht]
  \centering
  \reportfig{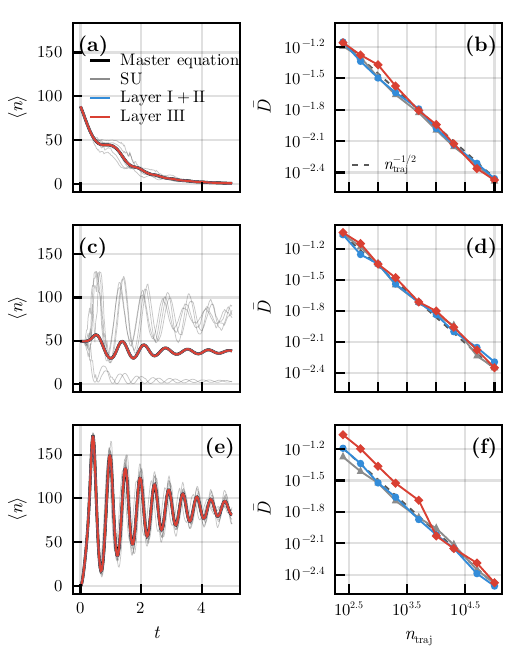}{\columnwidth}
  \caption{Accuracy and convergence for the Kerr resonator model. The rows from top to bottom refer to the
  dim, bistable and bright regimes, respectively.  Left column, panels \textbf{(a,c,e)}: mean photon number
  $\langle \hat n\rangle$ calculated from $10^5$ trajectories for standard unraveling (gray), Layer~I+II (light blue), and Layer~III (red), together with the 
  master-equation reference
  (black); six trajectories appear as thin gray curves, to highlight the different behavior between the mono- and bistable regimes.  Right
  column, panels \textbf{(b,d,f)}: averaged trace distance $\overline D$ from
  Eq.~\eqref{eq:trace-distance} against the number of trajectories
  $n_{\rm traj}$, for standard unraveling (gray triangles), Layer~I+II (light blue
  circles), and Layer~III (red 
  diamonds). For Layer~III, $m^{(-)}=20$, $m^{(+)}=60$ and $r_{\rm tol}=10^{-4}$ in all panels.  The dashed
  $1/\sqrt{n_{\rm traj}}$ line is a guide for the eye and represents the expected scaling of $\bar{D}_{\rm tr}$, which all methods follow.}
  \label{fig:kerr_correctness}
\end{figure}

First, in Fig.~\ref{fig:kerr_correctness} we stress test \mbox{DiSLOU} by comparing the accuracy of its Layers to the master-equation dynamics and standard unraveling trajectories. The initial states are far from the attractors of the system. This results in transient dynamics whose activity Layer~I struggles to suppress and which lead to frequent fallbacks in Layer~III.

For the dim regime in Fig.~\ref{fig:kerr_correctness}(a), we prepare a coherent state with many photons and observe the decay of its photon number to an near-vacuum steady state.
For the bistable regime in Fig.~\ref{fig:kerr_correctness}(c), the initial state is a coherent state with amplitude between those of the two attractors.
For the bright regime in Fig.~\ref{fig:kerr_correctness}(e), the initial state is close to the vacuum.
While single trajectories can significantly deviate from the result of the master equation, their average faithfully reproduces the results of the Liouvillian evolution.
In  Figs.~\ref{fig:kerr_correctness}(b, d, f) we study, as a function of the number $n_{\rm traj}$ of quantum trajectories, the convergence of the reconstructed density matrix $\hat\rho_{\rm MC}=\frac{1}{n_{\rm traj}}\sum_{r=1}^{n_{\rm traj}}\ketbra{\psi_r(t)}{\psi_r(t)}$ towards the density matrix $\hat\rho_{\rm ME}$ obtained via the master equation.
As a metric, we use the trace distance
\begin{equation}
\overline D=\frac1{N_t}\sum_{k=1}^{N_t}
\frac12\left\lVert\hat\rho_{\rm MC}(t_k)-\hat\rho_{\rm ME}(t_k)\right\lVert_1,
\label{eq:trace-distance}
\end{equation}
where we average the distance over $N_t=21$ equally spaced times.
As expected, evolutions under Layers~I and II agree with the results of the master equation and follow the expected scaling $\bar{D}\propto n_{\rm traj}^{-\frac12}$.
More interestingly, even the Layer~III evolution closely reproduces the results of the master equation within statistical uncertainty, including throughout the oscillatory transients generated by the deliberately mismatched initial states.
Strikingly, it is sufficient to consider as few as $m=20$ states to observe this agreement, compared to the $N = 190$ Hilbert space dimension required to obtain convergence using the full Hilbert space method.
The absence of a visible saturation with increasing $n_{\rm traj}$ indeed indicates that, within statistical resolution, the error remains dominated by Monte Carlo sampling rather than by the Layer~III projection.
Finally, let us remark that, in the bistable regime [Fig.~\ref{fig:kerr_correctness}(c)], individual trajectories also relax towards both metastable branches, confirming that the two gauges capture the regions explored by the dynamics.

While these calculations were tailored to provide a deliberately stringent test of \mbox{DiSLOU},
in App.~\ref{app:kerr_correctness_timings} we show that nonetheless \mbox{DiSLOU} reaches a fixed accuracy target faster than the standard unraveling method in all regimes.

\subsection{Performance}
\label{sec:kerr_performance}

\begin{figure}[!t]
  \centering
  \reportfig{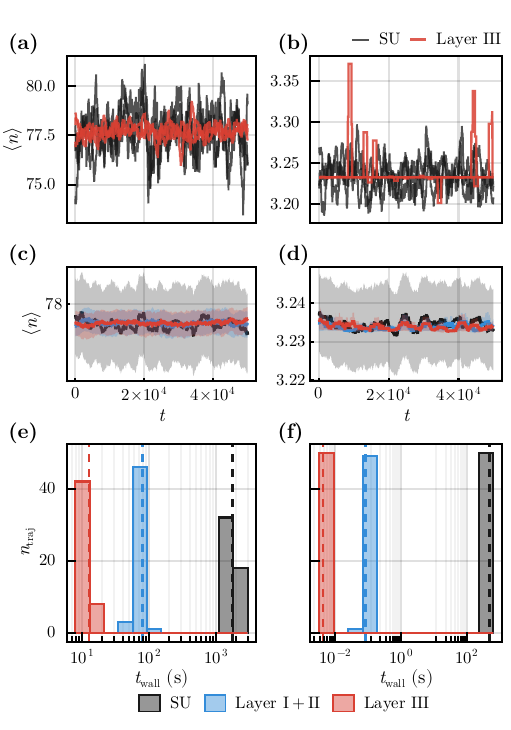}{\columnwidth}
  \caption{Benchmark on the Kerr resonator model. \textbf{(a,b)}~ Occupation $\langle \hat n \rangle$ for three standard unraveling (black) and three Layer~III
  (red) trajectories initialized in the high and low occupation regions,
  respectively: no large differences between methods are visible.  A 21-point moving average is used for clarity. 
  \textbf{(c,d)}~Ensemble mean (line) and one standard error (band) of $\langle \hat n\rangle$ for standard unraveling, Layer~I+II and Layer~III
  smoothed with a 51-point moving average.
  \textbf{(e,f)}~Wall-time histograms for the high and low occupation initial states respectively. $n_{\rm traj}$ refers to the number of trajectories for each method in each per-trajectory wall time ($t_{\rm wall}$) bin. 50 trajectories for each method are used, all evolved to
  $\kappa T=5\times10^4$. For Layer~III, $m^{(-)}=11$, $m^{(+)}=40$ and $r_{\rm tol}=10^{-4}$.  Dashed vertical lines mark medians. We find a significant reduction in wall time when using DiSLOU, compared to the standard unraveling.}
  \label{fig:kerr_performance}
\end{figure}

Having established accuracy, we turn to the regime \mbox{DiSLOU} is designed to accelerate: long trajectories that spend most of their time near a metastable attractor. In the bistable regime, we initialize coherent states at either the dim or bright semiclassical fixed point and evolve them up to $\kappa T=5\times10^4$. 
Figure~\ref{fig:kerr_performance} compares the trajectories evolved with \mbox{DiSLOU} and their propagation times with those of the standard unraveling method.
The computational gain is substantial, around one order of magnitude for Layer~I+II, and an additional order of magnitude for Layer~III, with no significant differences in the average dynamics of the observable.
Notice the difference in performance between trajectories that lie in the dim and bright branches.
The discrepancy follows from the bananization\footnote{ ``Kerr-type nonlinearities [\dots] tend to create a banana-shaped distortion of the coherent states in phase space, a phenomenon that is sometimes referred to as bananization''~\cite{blais2021}.} of the high-occupancy metastable state, which increases the residual activity in Eq.~\eqref{eq:gauge_rate} through the irreducible variance term, while its low-occupancy counterpart remains close to coherent, as can also be seen in the Wigner function plot of Fig.~\ref{fig:kerr_bistable}(b).

\section{Two-mode dissipative cat}
\label{sec:cat}

We consider a high-quality-factor resonator (the memory mode $a$) coupled to a strongly dissipative and driven auxiliary resonator (the buffer mode $b$) through a two-to-one photon-exchange interaction $g_2$ activated by a dedicated pumping mechanism. 
Following the derivation in~\cite{lescanne2020,reglade2024,berdou2023}, the Hamiltonian, in an appropriate rotating frame, reads
\begin{equation}
\begin{split}
\hat H&=\Delta_a \hat n_a+\Delta_b \hat n_b
-\frac{K_a}{2}\hat a^{\dagger 2}\hat a^2
-\frac{K_b}{2}\hat b^{\dagger 2}\hat b^2\\
&-\frac{\chi}{2}\hat n_a\hat n_b
+g_2\left(\hat a^2\hat b^\dagger+\hat a^{\dagger 2}\hat b\right)
+(\varepsilon_b^* \hat b+ \varepsilon_b \hat b^\dagger).
\end{split}
\label{eq:detuned_cat_hamiltonian}
\end{equation}
Here, $\Delta_a$ ($\Delta_b$) and $K_a$ ($K_b$) are the memory (buffer) detunings and self-Kerr nonlinearities, respectively. $\chi$ is the cross-Kerr coupling.
The jump operators are
\begin{align}
\hat C_1&=\sqrt{\kappa_1}\,\hat a,
&\hat C_b&=\sqrt{\kappa_b}\,\hat b,
&\hat C_\phi&=\sqrt{\kappa_\phi}\,\hat a^\dagger \hat a
\label{eq:cat_channels}
\end{align}
corresponding to memory photon loss, buffer photon loss, and memory dephasing channels, respectively.

In the absence of memory photon loss, dephasing, and Kerr nonlinearities, this system stabilizes the coherent states $\lvert\pm\alpha\rangle$~\cite{mirrahimi2014,leghtas2015,touzard2018}
as well as their quantum superpositions $\ket{\mathcal{C}^{\pm}_\alpha} \propto \ket{\alpha} \pm \ket{-\alpha}$ known as Schr\"odinger cat states. 
The manifold defined by their span can be used to define a bosonic encoding for a qubit~\cite{mirrahimi2014,guillaud2019,puri2020,guillaud2021,chamberland2022,leregent2023,gautier2023,gouzien2023,ruiz2025,labaymora2026chiralcat}.
The presence of weak single-photon loss and dephasing introduces errors. 
While bit-flips, which are connected to relaxation between the coherent states, are exponentially suppressed with photon number in the memory~\cite{minganti2016,lescanne2020,guillaud2019,berdou2023}, the phase-flip rate between cats grows linearly with it: this discrepancy can be exploited to produce noise-biased qubits. 
Protected storage, coherent control, and error correction on such systems have been demonstrated experimentally~\cite{lescanne2020,marquet2024autoparametric,reglade2024,marquet2024control,putterman2025coherence,putterman2025qec}.

\subsection{Ideal cat operated at resonance}

We consider an ideal setup, where $\Delta_a = \Delta_b = K_b = \chi =0$.
For consistency with current experimental efforts, we fix
$K_a/2\pi=-10\,{\rm kHz}$, $g_2/2\pi=\sqrt5\,{\rm MHz}$, 
$|\varepsilon_b|/2\pi=6\sqrt5\,{\rm MHz}$,
$\kappa_1/2\pi=10 \,{\rm kHz}$, $\kappa_b/2\pi=20 \,{\rm MHz}$, $\kappa_\phi/2\pi=1\, {\rm kHz}$.
We also choose the phase of $\varepsilon_b$ so that at the steady state $\operatorname{Tr}[\hat a^2 \hat\rho_{\rm ss}] = \alpha^2$ with $\alpha$ real and positive.
For this choice, the semiclassical approximation predicts that the metastable coherent states $\ket{\pm \alpha}$ have real $\alpha$, with~\cite{berdou2023}
$
\alpha^2\simeq6
$.

We use the prediction of the semiclassical approximation to choose the gauge. Since we now have $\lvert s\alpha\rangle$, with $s=\pm1$, we get
\begin{equation}
\zeta_1^{(s)}=-s\sqrt{\kappa_1}\alpha,
\qquad
\zeta_b^{(s)}=0,
\qquad
\zeta_\phi^{(s)}=-\sqrt{\kappa_\phi}|\alpha|^2.
\label{eq:cat-shifts}
\end{equation}
Notice that the dephasing jump operator also has a shifted gauge, identical for the two semiclassical attractors\footnote{Note that, even for a perfectly positioned coherent state $\ket{\alpha}$, the dephasing channel retains an irreducible residual activity $A(\ket{\alpha})=\Var_{\ket{\alpha}}(\sqrt{\kappa_\phi}\hat n)=\kappa_\phi\lvert\alpha\rvert^2$.}. 
We leave the buffer channel undisplaced since the buffer almost always remains close to vacuum in the regime $\kappa_b\gg4|\alpha g_2|$ considered here~\cite{gautier2023,Ferrari2026BitFlips}.

\subsubsection{Slow quadrature relaxation}
\label{sec:cat_quadrature}

\begin{figure}[!htbp]
\centering
\reportfig{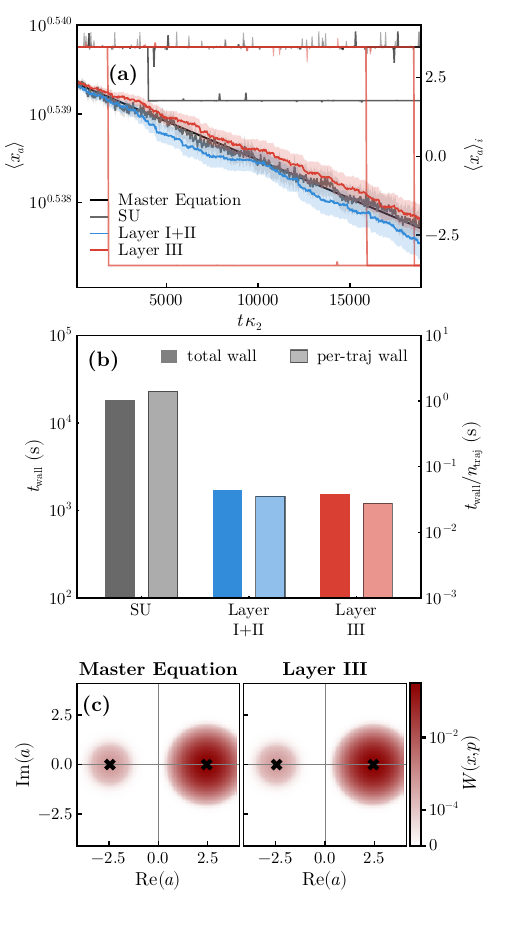}{\columnwidth}
\caption{Relaxation towards the steady state of the resonant cat
of Eqs.~\eqref{eq:detuned_cat_hamiltonian},~\eqref{eq:cat_channels}.
\textbf{(a)}~Average $\langle \hat x_a(t)\rangle$ on the left logarithmic axis for the master-equation reference (black), standard unraveling
(gray), Layer I+II (light blue), and Layer III (red).  Shaded bands show
one standard error. The fitted values of $\Gamma_{\rm decay}$ are $(1.91\pm0.41)\times10^{-7}\,\rm \kappa_2$, $(2.05\pm0.42)\times10^{-7}\,\rm \kappa_2$ and $(1.91\pm0.42)\times10^{-7}\,\rm \kappa_2$ for standard unraveling, Layer~I+II and Layer~III respectively, with 13100, 48050 and 54700 trajectories in each ensemble and $\kappa_2\equiv\frac{4\vert g_2\rvert^2}{\kappa_b}=1\,\rm MHz$. Single standard unraveling (gray) and Layer~III (red) trajectories exhibiting large phase space excursions as indicated by the right-hand axis.
\textbf{(b)}~Propagation wall time for the three trajectory methods.
Dark bars show the total time to reach 20\% precision on the left axis; light bars show the mean per
completed trajectory on the right axis. DiSLOU resolves the decay rate $\Gamma_{\rm decay}$ in around one order of magnitude less time compared to the standard unraveling.
\textbf{(c)}~Final memory Wigner functions at $t\kappa_2=2\times10^4$ from the
master equation and the averaged
Layer~III trajectories.  The
Wigner maps share a symlog color scale; crosses mark the stable semiclassical solutions. In all panels, the Fock cutoffs for the memory and buffer modes are respectively $N_a=30$, $N_b=8$, for Layer~III we use $m^{(s)}=100$, $r_{\rm tol}=10^{-4}$.}
\label{fig:cat_quadrature}
\end{figure}

In Fig.~\ref{fig:cat_quadrature} we initialize the system in
$\lvert\psi_0\rangle=\lvert\alpha\rangle_a\otimes\lvert0\rangle_b$
and evolve the full two-mode model of Eqs.~\eqref{eq:detuned_cat_hamiltonian} and~\eqref{eq:cat_channels} up to $\kappa_2t=2\times10^4$. 
To characterize the relaxation rate of the metastable lobes $\ket{\pm \alpha}$, which is associated with bit-flip errors, in Fig.~\ref{fig:cat_quadrature}(a) we plot the expectation value of
\begin{equation}
\hat x_a=\frac{\hat a+\hat a^\dagger}{\sqrt2}
\label{eq:cat-quadrature}
\end{equation}
obtained with standard unraveling trajectories and the \mbox{DiSLOU} approach.
The average expectation value agrees for all methods and shows a decay compatible with 
\begin{equation}
\langle \hat x_a(t)\rangle
=x_0 e^{-\Gamma_{\rm decay}t},
\label{eq:cat-fit}
\end{equation}
where $x_0$ is the initial amplitude and $\Gamma_{\rm decay}$ is the decay rate.

Even though the average over trajectories is independent of the unravelings, the dynamics of individual trajectories is not, as shown in Fig.~\ref{fig:cat_quadrature}(a). 
Standard unraveling trajectories are \textit{qualitatively} different from \mbox{DiSLOU} ones. 
For the former, $x_a$ explores intermediate values between $\pm \sqrt{2} \alpha$, while \mbox{DiSLOU} trajectories show an abrupt switch between the coherent states of opposite amplitude.
The origin of this difference follows directly from the measurement records: for standard memory photon loss, the system is pushed towards cat-like states, in which photon-loss events induce shifts in parity~\cite{bartolo2017,Ferrari2026BitFlips,Ferrari2026Erasures}.
The trajectories obtained with the \mbox{DiSLOU} method, instead, more closely resemble homodyne trajectories, and the optimal gauge maximally distinguishes the $\lvert+\alpha\rangle$ and $\lvert-\alpha\rangle$ lobes.

Different unravelings also produce different sampling variances for a given observable.
It follows that any computational advantage cannot be assessed from the cost of a single trajectory alone. 
We thus compare the total computational effort required to estimate $\Gamma_{\rm decay}$ at a fixed statistical precision.
Specifically, for each method we simulate a number of trajectories high enough to reach a fitted decay rate with relative precision of $20\%$ with respect to the master-equation reference.
We then repeat this procedure several times.
Figure~\ref{fig:cat_quadrature}(b) reports the accumulated propagation time required to reach this target and the average cost of a single trajectory. The resulting ensembles contain approximately $10^5$ trajectories for the standard algorithm and around $5 \cdot 10^5$ for \mbox{DiSLOU}.\footnote{Other variables, such as photon number and parity, show the opposite scaling: \mbox{DiSLOU}, resembling homodyne trajectories, displays lower variance than the standard unraveling algorithm.}
Despite requiring substantially more realizations, both \mbox{DiSLOU} variants reach the target precision in more than an order of magnitude less total propagation time than standard unraveling.
Indeed, the \mbox{DiSLOU} trajectories use on average less than $2\%$ of the time required by the corresponding standard unraveling trajectory.

Finally, in Fig.~\ref{fig:cat_quadrature}(c) we plot the Wigner functions of the final reconstructed density matrix of the memory obtained from both the master equation and \mbox{DiSLOU}.
No sizable differences are visible.

\subsection{Non-ideal cat out of resonance}
\label{sec:non-ideal-cat}

Now we consider
$K_a/2\pi = -80\,{\rm kHz}$,  $K_b/2\pi =-1.20\,{\rm MHz}$,
$\chi/2\pi =-600\,{\rm kHz}$, $g_2/2\pi =1.34\,{\rm MHz}$, $\kappa_1/2\pi =5\,{\rm kHz}$, $\kappa_b/2\pi =15.0\,{\rm MHz}$, and $|\varepsilon_b|/2\pi =7.60\,{\rm MHz}$.
We retain photon loss from the memory and buffer modes, but we do not account for dephasing in this example\footnote{In this case, dephasing would allow faster transition rates between the dim and the bright phases, de facto simplifying the problem.}.

When calibrating cat systems, one often investigates the response of the system to effective detunings~\cite{berdou2023}.
As in the case of the single-mode Kerr resonator, detunings can give rise to additional multistability phenomena.
Here we fix $\Delta_a/2\pi =-500\,{\rm kHz}$  and $\Delta_b/2\pi =41.0\,{\rm MHz}$, enabling the existence of three metastable states. 
We label them by $g\in\{l,-,+\}$, corresponding to the low-occupation state and the two high-occupation states, respectively. Writing
$\alpha=\langle \hat a\rangle$ and $\beta=\langle \hat b\rangle$, their amplitudes are approximately
\begin{align}
(\alpha_l,\beta_l)
&\simeq
\left(0,-0.18-0.03i\right),
\nonumber\\
(\alpha_\pm,\beta_\pm)
&\simeq
\left(\pm(4.85+3.28i),-0.35-0.87i\right).
\label{eq:detuned_cat_fixed_points}
\end{align}
The two high-occupation solutions are related by the $\hat a\rightarrow-\hat a$ symmetry and have an appreciably occupied buffer, whereas the low-occupation solution lies close to the memory vacuum and has only a small buffer amplitude.

\begin{figure}[!htb]
  \centering
  \reportfig{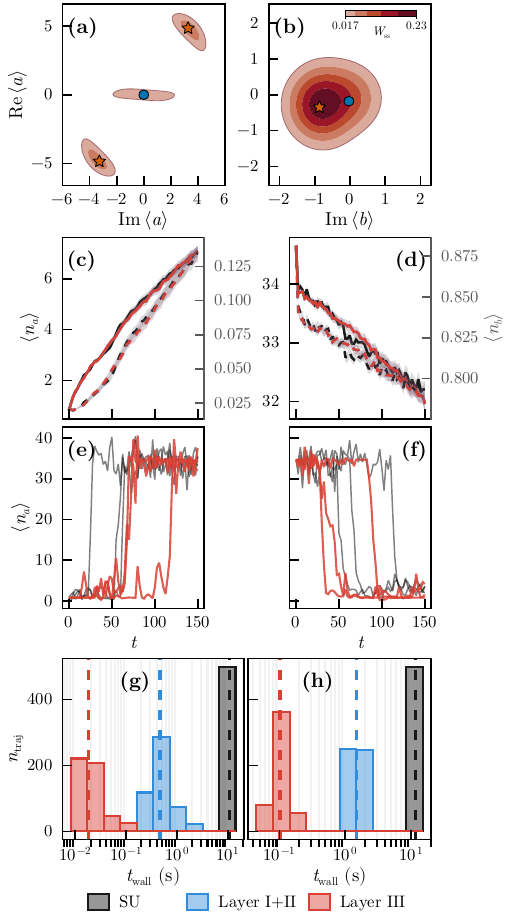}{\columnwidth}
  \caption{Benchmark on the detuned two-mode cat model.
  \textbf{(a,b)}: Wigner functions of the steady state for memory and buffer mode, respectively. Orange (blue) markers indicate the high (low) occupation semiclassical solutions.
  \textbf{(c,d)}: ensemble means over 3000 trajectories of the memory occupation $\langle \hat n_a\rangle$ and 
  buffer occupation $\langle \hat n_b\rangle$ vs time. Gray,and red denote
  standard unraveling and Layer~III propagation respectively.
  \textbf{(e,f)}: Three $\langle
  \hat n_a\rangle$ traces for single standard unraveling and Layer~III trajectories, chosen to illustrate switching between the low and high occupation regions.  
  \textbf{(g,h)}: Wall-time histograms. $n_{\rm traj}$ refers to the number of trajectories for each method in each per-trajectory wall time ($t_{\rm wall}$) bin. Dashed vertical lines indicate the per-method median. 
  Panels (c,e,g) use initial states in the low-occupation region, panels (d,f,h) in the high-occupation one. In both cases, we find a significant reduction in the median wall time for DiSLOU compared to the standard unravelling. For Layer~III, we use $m^{(l)}=200$, $m^{(\pm)}=300$ and $r_{\rm tol}=10^{-4}$.}
  \label{fig:detuned_transfer}
\end{figure}

Quantum mechanically, these fixed points give rise to three metastable regions, which we denote by $\ket{\alpha_l}$ and $\ket{\psi_\pm}$: these can be identified in the steady-state Wigner functions plotted in panels (a,b) of Fig. ~\ref{fig:detuned_transfer}.
The corresponding states are not coherent: squeezing and higher-order fluctuations generally imply
$\langle \hat n_a\rangle_{\psi_g}\neq|\alpha_g|^2$ and
$\langle \hat n_b\rangle_{\psi_g}\neq|\beta_g|^2$, with nonzero fluctuations of both modes: these introduce jumps that are impossible to reduce by displacement.

This case is also one where the master equation requires very significant numerical resources, especially if one has to resolve the various transition rates of the system: to reach convergence, the memory and buffer Fock spaces need to be truncated at $N_a=85$ and $N_b=16$ respectively, giving a total Hilbert-space dimension $N=N_aN_b=1360$.

We construct one gauge for each of the three semiclassical fixed points identified above. 
Displacing the buffer channel is essential in this regime. In contrast with the near-ideal resonant cat of Sec.~\ref{sec:cat}, the two high-occupation metastable states have a finite buffer amplitude $\beta_\pm$. Since $\kappa_b/\kappa_1=3000$, leaving this channel undisplaced would make buffer photon loss dominate the event activity. 
For $g\in\{l,-,+\}$, the memory- and buffer-channel shifts are
\begin{equation}
\boldsymbol{\zeta}^{(g)}
=
\left(
-\sqrt{\kappa_1}\,\alpha_g,\,
-\sqrt{\kappa_b}\,\beta_g
\right),
\label{eq:detuned_cat_output_centres}
\end{equation}
where $\alpha_g$ and $\beta_g$ are the results of the semiclassical analysis,
so that 
$
\hat C_a^{(g)}=\sqrt{\kappa_1}\left(\hat a-\alpha_g\right),\,
\hat C_b^{(g)}=\sqrt{\kappa_b}\left(\hat b-\beta_g\right).
$

We benchmark \mbox{DiSLOU} starting from each of the three metastable regions identified above. For each gauge $g\in\{l,-,+\}$, we initialize the dynamics in the normalized eigenmode of $\hat H_{\rm eff}^{(g)}$ with the smallest eigenvalue decay rate. This choice places the initial state in the longest-lived region supported by the corresponding locally optimal gauge.

Fig.~\ref{fig:detuned_transfer}(c,d) compares averaged expectation values obtained from $3000$ standard unraveling, Layers~I+II, and Layer~III trajectories for the low-occupation and one of the high-occupation initial conditions. The memory and buffer occupations, $\langle \hat n_a\rangle$ and $\langle \hat n_b\rangle$, agree across the three methods within statistical uncertainty throughout the simulated evolution. \mbox{DiSLOU} therefore reproduces the ensemble dynamics not only near each metastable region, but also during the switching processes connecting them.

Panels (e,f) display these transitions at the level of individual trajectories. Representative standard unraveling and Layers~I+II trajectories exhibit large occupation excursions and switches between the high and low occupation metastable regions. The displaced unraveling therefore continues to correctly resolve the dynamics when three gauges are required, rather than the two in the previous examples.

The computational advantage is shown by  Fig.~\ref{fig:detuned_transfer}(g,h).
Across the different initial states, the median propagation times of \mbox{DiSLOU} are smaller than those of standard unraveling by two to three orders of magnitude, with a significant speedup achieved both by reducing the activity and the evolution space.

We conclude that the locally optimal construction extends naturally to more than two metastable regions, whilst all layers remain advantageous even at substantially larger Hilbert-space dimensions.

\begin{figure}[!htb]
  \centering
  \reportfig{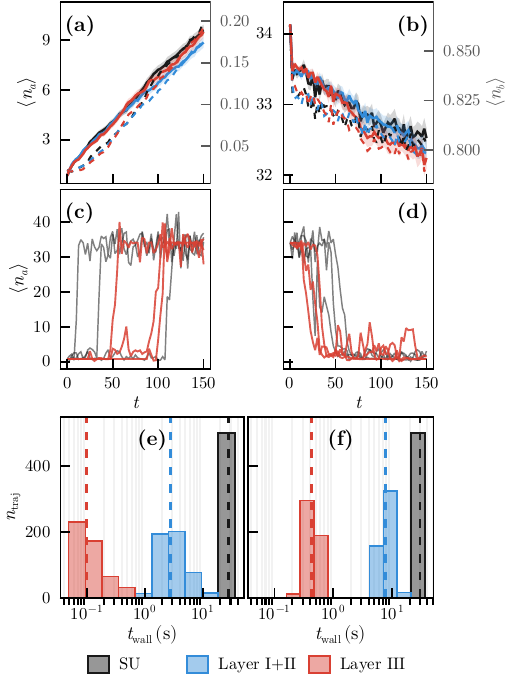}{\columnwidth}
  \caption{Benchmark on the detuned two-mode cat model whose memory is coupled to a two-level system.
  \textbf{(a,b)}: ensemble means over 1600 trajectories of the memory occupation $\langle \hat n_a\rangle$ and 
  buffer occupation $\langle \hat n_b\rangle$ vs time. Gray, blue, and red denote
  standard unraveling, Layer I+II, and Layer~III propagation respectively.
  \textbf{(c,d)}: Three $\langle
  \hat n_a\rangle$ values for single standard unraveling and Layer~III trajectories, chosen to illustrate switching between the low and high occupation regions.  
  \textbf{(e,f)}: Wall-time histograms. $n_{\rm traj}$ refers to the number of trajectories for each method in each per-trajectory wall time ($t_{\rm wall}$) bin.
  Panels (a,c,e) use initial states in the low-occupation region, panel (b,d,f) in the high-occupation one. The performance advantage of DiSLOU persists in both regimes for this larger system. For Layer~III, we use $m^{(l)}=400$, $m^{(\pm)}=600$ and $r_{\rm tol}=10^{-4}$.}
  \label{fig:tls_ancilla_benchmark}
\end{figure}

\subsection{Coupling to a two-level system}
\label{sec:tls}

To verify the functioning of \mbox{DiSLOU} on regimes that are otherwise difficult to access, we increase the size and complexity of the model by strongly coupling the memory
mode of the previous section to a detuned, long-lived two-level system (TLS), such as those commonly observed in superconducting circuit experiments~\cite{simmonds2004,shalibo2010,sarabi2015,muller2019tls}.
In many cases, their effects cannot be traced out, as it would result in memory effects in the master equation, preventing the derivation of a Lindblad equation and thus the use of standard quantum trajectories.
The Hamiltonian becomes
\begin{equation}
\begin{split}
\hat H={}&(\Delta_a+\lambda)\hat n_a+\Delta_b \hat n_b
-\frac{K_a}{2}\hat a^{\dagger2}\hat a^2
-\frac{K_b}{2}\hat b^{\dagger2}\hat b^2\\
&-\frac{\chi}{2}\hat n_a\hat n_b
+g_2\left(\hat a^2\hat b^\dagger+\hat a^{\dagger2}\hat b\right)
+(\varepsilon_b^*\hat b+\varepsilon_b \hat b^\dagger)\\
&+\Delta_q\hat\sigma^+\hat\sigma^-
+g_q\left(\hat a\hat\sigma^++\hat a^\dagger\hat\sigma^-\right),
\end{split}
\label{eq:tls_cat_hamiltonian}
\end{equation}
where $\lambda=g_q^2/\Delta_q$ compensates the shift in the memory frequency induced by the dispersive coupling, $\Delta_q$ is the TLS detuning, $g_q$ its coupling strength to the memory and $\hat\sigma^{\pm}$ the two-level raising/lowering operators. We retain the parameters of the detuned cat of Sec.~\ref{sec:non-ideal-cat} and additionally take
$\Delta_q/2\pi=300\,{\rm MHz}$ and $g_q/2\pi=10\,{\rm MHz}$, with
$\lambda/2\pi=0.33\,{\rm MHz}$ and
$(\Delta_a+\lambda)/2\pi=-0.167\,{\rm MHz}$.
A jump operator related to the TLS is also introduced as
$\hat C_q=\sqrt{\kappa_q}\hat\sigma^-$, with $\kappa_q/2\pi=10\,\rm kHz$.

With cutoffs $N_a=85$ and $N_b=16$, the TLS increases the Hilbert-space
dimension to $N=2720$, making evolution with the master equation
impractical. We use the same three-gauge construction as in Eq.~\eqref{eq:detuned_cat_output_centres} for the dissipation channels of the memory and buffer modes, while the TLS jump operator is left undisplaced. The semiclassical equations for $\alpha_g$, $\beta_g$ are solved also taking into account the TLS.

Figure~\ref{fig:tls_ancilla_benchmark} shows that Layers~I+II and Layer~III reproduce
the ensemble dynamics and the switching events obtained with the standard unraveling.
At the same time, \mbox{DiSLOU} retains a substantial reduction in propagation
time, visible in particular when Layer~III is enabled.

\section{Discussion, limitations, and perspectives}
\label{sec:conclusion}

In this work, we have introduced \mbox{DiSLOU}, which combines locally optimal unravelings, diagonal propagation, and local reduced eigenspaces to simulate quantum trajectories in metastable open quantum systems.
An implementation of \mbox{DiSLOU} in Julia is available as an open-source package~\cite{DiSLOU_git}.

The advantage of \mbox{DiSLOU} is based on the use if three layers.
Layers~I and II are exact unravelings of the Lindblad equation, while Layer~III introduces a controlled projection with an automatic fallback to the full space. On the driven-dissipative Kerr resonator and on two-mode dissipative cats with up to three metastable regions and Hilbert-space dimensions up to $N=2720$, \mbox{DiSLOU} agrees with master-equation or standard-trajectory references within statistical uncertainty. At the same time, it reduces the computational cost by one to several orders of magnitude, even when, as for the bit-flip rate of the resonant cat, it needs more trajectories to reach a given precision.

The method is best suited to dynamics with moderate fluctuations, where the reduction in quantum jumps is effective except for a limited number of rare events. Little advantage should be expected for short evolutions, dynamics that explore a large part of Hilbert space for a significant proportion of time, or systems for which diagonalizing the effective Hamiltonian is prohibitive. Furthermore, numerical stability requires the eigenvectors of the effective Hamiltonian to be sufficiently well conditioned. This regime could be broadened by a more stable spectral representation, for example a Schur-based formulation, which would retain the first-passage construction for strongly nonnormal effective Hamiltonians.

Between jumps, the propagation consists predominantly of dense matrix-vector products and quadratic forms, without the adaptive ODE stepping of standard unraveling, which makes \mbox{DiSLOU} well suited to hardware accelerators. Our implementation already runs each trajectory on GPU arrays (see App.~\ref{app:computational-platform}), and the batched evolution of large trajectory ensembles provides a natural route towards larger Hilbert spaces.

Finally, \mbox{DiSLOU} could be extended to time-dependent Floquet problems, a setting recently studied in the context of dissipative phase transitions~\cite{Mercurio2026FloquetDPT}. In the regime where a dominant carrier frequency defines the rotating frame and smaller time-dependent terms act as perturbations, one possible approach is a pseudomode decomposition involving a Floquet non-Hermitian Hamiltonian.

\section*{Acknowledgements}
\label{sec:acknowledgements}
The authors acknowledge useful discussions with Filippo Ferrari, Adrià Labay-Mora, Philippe Campagne-Ibarcq and with colleagues in the various experimental and theoretical departments of Alice \& Bob, in particular Giulio Campanaro, Nicolas Didier, Th\'eo Malas-Danz\'e, Andr\'e Melo, Nora Reini\'c, and David Schlegel.

\section*{Author contributions}
F.M. and P.P. conceived the project. P.P. developed the code and performed the simulations, with input from A.M.\ P.P. and A.M. prepared and released the \code{DiSLOUTrajectories.jl} Julia package. F.M. and P.W. supervised the project. All authors discussed the results and contributed to the manuscript.

Large language models (LLMs), including GPT-6 Astra (OpenAI) and Claude Fable 5.1 (Anthropic), were employed to assist with code development, figure preparation, and proofreading of the manuscript. The original ideas behind the method are the authors' own. All the material was reviewed by the authors, who take full responsibility for the content of this work.

\newpage

\appendix 

\numberwithin{equation}{section}
\counterwithin{figure}{section}

\section{Automatic construction of the gauge set}
\label{app:gauge-searching}

The finite gauge set of Layer~I can be constructed automatically without prior knowledge of the metastable states. We consider here two complementary procedures: evaluating and solving the semiclassical equations associated with the Hamiltonian and jump operators, or identifying the long-lived regions from a preliminary ensemble of short quantum trajectories.

\subsection{Semiclassical fixed-point search}

For the bosonic models considered here, the semiclassical equations can be evaluated directly from the Hamiltonian and jump operators. Let $\hat a_j$ denote the annihilation operators and $\alpha_j=\langle \hat a_j\rangle$. Starting from Eq.~\eqref{eq:lindblad}, the adjoint Lindblad equation gives
\begin{equation}
\dot\alpha_j
=
\left\langle
i[\hat H,\hat a_j]
+\sum_\mu
\left(
\hat C_\mu^\dagger \hat a_j \hat C_\mu
-\frac12\{\hat C_\mu^\dagger \hat C_\mu,\hat a_j\}
\right)
\right\rangle .
\label{eq:app-mean-field-general}
\end{equation}
We construct the matrix of the operator on the right-hand side once for each mode and evaluate its expectation value in the coherent product state $\ket{\boldsymbol{\alpha}}=\bigotimes_j\ket{\alpha_j}$.
For normally ordered products, these expectation values factorize in terms of the amplitudes $\alpha_j$ and their complex conjugates, yielding the first-order mean-field equations
$
\dot\alpha_j=
F_j(\boldsymbol{\alpha},\boldsymbol{\alpha}^{*})
$.

The fixed points are the simultaneous roots
\begin{equation}
F_j(\boldsymbol{\alpha}^{(g)},
\boldsymbol{\alpha}^{(g)*})=0
\qquad \forall j .
\label{eq:app-mean-field-roots}
\end{equation}
Numerically, we rewrite the complex amplitudes as the real phase-space vector
\begin{equation}
\boldsymbol{x}
=
(\Re\alpha_1,\ldots,\Re\alpha_M,
\Im\alpha_1,\ldots,\Im\alpha_M)\,,
\end{equation}
and similarly map the $F_j$ to a real residual vector
$\boldsymbol{R}(\boldsymbol{x})$. We sample a bounded region of phase space with a regular grid of initial guesses and refine each candidate using a trust-region method for the nonlinear system $\boldsymbol{R}(\boldsymbol{x})=0$~\cite{pal2025nonlinearsolve}. The Jacobian
$J=\partial\boldsymbol{R}/\partial\boldsymbol{x}$ is evaluated using forward-mode automatic differentiation. Converged roots within the prescribed occupation bounds are retained and subsequently deduplicated.

The same Jacobian determines their local stability. We retain a fixed point
when
\begin{equation}
\max_\ell\Re\lambda_\ell
\!\left[J(\boldsymbol{x}^{(g)})\right]<0 ,
\label{eq:app-mean-field-stability}
\end{equation}
with a small numerical margin below zero. Finally, evaluating the
jump-operator expectations in the coherent product states at the stable fixed points gives
\begin{equation}
\zeta_\mu^{(g)}
=
-\bra{\boldsymbol{\alpha}^{(g)}}\hat C_\mu\ket{\boldsymbol{\alpha}^{(g)}}.
\label{eq:app-mean-field-gauges}
\end{equation}
For instance, $\hat C_\mu=\sqrt{\kappa_j}\hat a_j$ gives
$\zeta_\mu^{(g)}=-\sqrt{\kappa_j}\alpha_j^{(g)}$.

\subsection{Preliminary trajectories and clustering}

A second route determines the gauge set directly from the quantum dynamics.
We initialize short trajectories from coherent product states distributed over
the relevant region of phase space and propagate them using standard quantum-jump dynamics~\cite{mercurio2025}. For each realization $r$, we characterize the
region reached after the initial relaxation by averaging the mode amplitudes
over a terminal time window,
\begin{equation}
\overline{\alpha}_j^{(r)}
=
\frac{1}{\Delta T}
\int_{T-\Delta T}^{T}
\dd t\,
\langle \hat a_j\rangle_{\psi_r(t)} .
\label{eq:app-cluster-amplitudes}
\end{equation}
Each trajectory is therefore represented by one point
$\overline{\boldsymbol{\alpha}}^{(r)}$ in the joint phase space of the modes.

We partition the terminal phase-space points using the density-based spatial clustering algorithm DBSCAN~\cite{ester1996dbscan,schubert2017dbscan}. Points whose neighborhood contains a minimum number of points, including themselves, form the cores of clusters. Connected core points belong to the same cluster; other points join the cluster of their nearest core point within the neighborhood radius, or are classified as noise if none is found. This construction is well suited to the present problem because the number of metastable regions does not need to be specified in advance and trajectories that have not relaxed into a well-populated region do not generate additional clusters. We additionally discard clusters containing less than a given fraction of the initial ensemble.

For each retained cluster $\mathcal I_g$, its phase-space position is computed
as the mean of its terminal points. The components of the gauges are obtained
from the expectation values of the unshifted jump operators averaged over the same trajectories and time
window,
\begin{equation}
\zeta_\mu^{(g)}
=
-\frac{1}{|\mathcal I_g|}
\sum_{r\in\mathcal I_g}
\frac{1}{\Delta T}
\int_{T-\Delta T}^{T}
\dd t\,
\langle \hat C_\mu\rangle_{\psi_r(t)} .
\label{eq:app-cluster-gauge}
\end{equation}

\section{Root solve for the quantum-jump time}
\label{app:root-solve}

We describe here how the first-passage time of Eq.~\eqref{eq:first_passage} is determined. During each no-jump interval the gauge is fixed, and we suppress its label. Starting from a normalized state, $s(0)=1$, and a random threshold $u\in[0,1)$, the next quantum jump occurs at the first time $\tau$ such that $s(\tau)=u$.

The survival probability is nonincreasing since
\begin{equation}
-\dot s(t)=
\sum_\mu
\bra{\widetilde\psi(t)}
\hat C_\mu^\dagger \hat C_\mu
\ket{\widetilde\psi(t)}
\geq 0 ,
\label{eq:survival_derivative}
\end{equation}
where $\hat C_\mu$ denotes the jump operators of the active gauge.

In our implementation, the diagonal propagation of Sec.~\ref{sec:full_spectral} is formulated as an exact integrator for the ordinary differential equation solvers used by \code{mcsolve} in \code{QuantumToolbox.jl}~\cite{mercurio2025,rackauckas2017differentialequations}. The jump times are therefore located by the same event-detection procedure as for standard quantum trajectories, but on the exact no-jump solution. Within each step, of duration $T$, the integrator provides an exact continuous output, $\ket{\widetilde\psi(t+\theta)}=VD(\theta)c$ for $0\leq\theta\leq T$, where $c$ are the coordinates at the beginning of the step.

At the end of each step we evaluate $s(t+T)$. If $s(t+T)>u$, monotonicity guarantees that no jump occurred within the step, so that $s$ never needs to be evaluated inside a step without a jump. Otherwise, the first passage is bracketed between $t$ and $t+T$, and the bracket is refined on the continuous output by the bracketing root solver used for event detection in \code{DifferentialEquations.jl}, a modified Anderson--Bj\"orck method~\cite{ganchovski2026}, which combines the superlinear convergence of secant-type iterations with the robustness of a bracket. Each evaluation of $s$ costs $O(N^2)$ in the full space and $O(m^2)$ in the reduced spaces of Layer~III (see Sec.~\ref{sec:reduced}), independently of the duration of the no-jump interval, and a few evaluations are needed per jump. At the crossing, the jump channel $\mu$ is drawn with probability proportional to $\lVert\hat C_\mu\widetilde\psi(\tau)\rVert^2$, the jump is applied, and the integration restarts from the normalized post-jump state, in the gauge and basis selected by Layers~I and III.

\section{Reduced-space update at quantum jumps}
\label{app:reduced-jump}

We provide further details on the direct treatment of quantum jumps between the reduced eigenspaces of Layer~III. For gauge $g$, let $V_m^{(g)}$ collect the $m$ retained right eigenvectors of $\Heff^{(g)}$, with the thin QR decomposition of Eq.~\eqref{eq:reduced-qr}. The orthogonal projector onto $\mathcal K_m^{(g)}$ is then $P_m^{(g)}=Q_m^{(g)}Q_m^{(g)\dagger}$, equivalent to the expression in Eq.~\eqref{eq:residual}, and a state $\ket{\psi}=V_m^{(g)}c$ in $\mathcal K_m^{(g)}$ is represented by its orthonormal coordinates $q=Q_m^{(g)\dagger}\ket{\psi}=R_m^{(g)}c$, with $\lVert q\rVert=\lVert\psi\rVert$.

Consider a jump in channel $\mu$ from a state represented by the coordinates $q^-$ in $\mathcal K_m^{(g)}$, and suppose that the following no-jump interval is assigned to gauge $g'$ and reduced space $\mathcal K_{m'}^{(g')}$. Before normalization, the jump produces
\begin{equation}
\ket{\phi}=X_\mu^{(g)}q^-,
\qquad
X_\mu^{(g)}=\hat C_\mu^{(g)}Q_m^{(g)} .
\end{equation}
Its projection onto the destination reduced space is the least-squares approximation $V_{m'}^{(g')}c^+\simeq\ket{\phi}$. Using Eq.~\eqref{eq:reduced-qr}, its orthonormal and eigenvector coordinates follow directly as
\begin{equation}
q^+=Q_{m'}^{(g')\dagger}X_\mu^{(g)}q^-,
\qquad
c^+=\left[R_{m'}^{(g')}\right]^{-1}q^+,
\label{eq:reduced-jump-transfer}
\end{equation}
where the $m'\times m$ matrix $Q_{m'}^{(g')\dagger}X_\mu^{(g)}$ is precomputed, and the action of $\left[R_{m'}^{(g')}\right]^{-1}$ is implemented as a numerically efficient upper-triangular solve. Since $c^-=(R_m^{(g)})^{-1}q^-$, Eq.~\eqref{eq:reduced-jump-transfer} is equivalent to $c^+=J_{\mu,\mathrm{red}}^{g'\leftarrow g}c^-$, with the map of Eq.~\eqref{eq:reduced-transfer}. The post-jump state is then normalized by normalizing $q^+$, as $Q_{m'}^{(g')}$ has orthonormal columns. Thus, the quantum jump, projection, and change of gauge are combined into a single map from the coordinates of one reduced eigenspace to those of the next.

All the other quantities required at the jump are quadratic forms in $q^-$ with precomputed $m\times m$ matrices. The jump channel is drawn with probabilities proportional to
\begin{equation}
\lVert\phi\rVert^2=\lVert\hat C_\mu^{(g)}\psi\rVert^2=(q^-)^\dagger X_\mu^{(g)\dagger}X_\mu^{(g)}q^- .
\end{equation}
The activities of Eq.~\eqref{eq:gauge_rate} after the jump, which select the gauge $g'$ through Eq.~\eqref{eq:hysteresis}, read
\begin{align}
A_{g'}\!\left(\frac{\phi}{\lVert\phi\rVert}\right)&=
\frac{(q^-)^\dagger M_\mu^{(g\to g')}q^-}{\lVert\phi\rVert^2},\\
M_\mu^{(g\to g')}&=\sum_\nu X_\mu^{(g)\dagger}\hat C_\nu^{(g')\dagger}\hat C_\nu^{(g')}X_\mu^{(g)} .
\end{align}
The accuracy of the projection is determined before accepting the new reduced space. Its relative residual is
\begin{equation}
r=\frac{\left\lVert\left(\mathbbm{1}-P_{m'}^{(g')}\right)\ket{\phi}\right\rVert}{\lVert\phi\rVert}\,.
\label{eq:reduced-jump-residual}
\end{equation}
Instead of constructing the $N$-dimensional objects of Eq.~\eqref{eq:reduced-jump-residual}, we precompute and cache the $m\times m$ matrices
\begin{equation}
B_\mu^{(g\to g')}=Y^\dagger Y,
\qquad
Y=\left(\mathbbm{1}-P_{m'}^{(g')}\right)X_\mu^{(g)},
\end{equation}
so that
\begin{equation}
r^2=\frac{(q^-)^\dagger B_\mu^{(g\to g')}q^-}{\lVert\phi\rVert^2}\,,
\end{equation}
and we can perform the residual check directly in reduced coordinates. Computing $Y=X_\mu^{(g)}-Q_{m'}^{(g')}\big(Q_{m'}^{(g')\dagger}X_\mu^{(g)}\big)$ before forming $Y^\dagger Y$ preserves the significant digits of small residuals.
If $r$ satisfies the tolerance of Eq.~\eqref{eq:residual}, the next interval starts directly from the coordinates in Eq.~\eqref{eq:reduced-jump-transfer}, and the whole jump costs $O(m^2)$ operations per channel and gauge. Otherwise, the $N$-dimensional state $\ket{\phi}$ is constructed and the full-space fallback of Sec.~\ref{sec:reduced} is used.

 \section{Computational platform and timing conventions} \label{app:computational-platform} 
 
 All simulations were performed on an Intel Xeon Platinum 8481C CPU at $2.70\,\mathrm{GHz}$, using four \mbox{OpenBLAS}~\cite{zhang2012openblas,wang2013augem} threads. Unless otherwise stated, timing benchmarks run trajectories serially, and exclude the one-time preparation cost of \mbox{DiSLOU}. Standard unraveling calculations use \code{mcsolve} from \code{QuantumToolbox.jl}~\cite{mercurio2025} v0.47.2 with Julia~\cite{Julia-2017} v1.12.6. 
 
 For standard unraveling, \code{mcsolve} operates on sparse \texttt{CSC} matrices, and its propagation times are essentially insensitive to the number of \texttt{BLAS} threads. \mbox{DiSLOU} instead performs dense linear-algebra operations and, for sufficiently large Hilbert spaces, its propagation performance scales approximately linearly with the number of \texttt{BLAS} threads. Its one-time preparation consists of one dense eigendecomposition and QR decomposition per gauge, at cost $O(N^3)$, shared by all trajectories.

The implementation keeps the array types of the input operators, so that it also runs on GPUs without any change. For instance, with \code{CUDA.jl}, it is sufficient to move the operators and the initial state of the example in Sec.~\ref{sec:hierarchy} to the GPU:
\begin{lstlisting}
using CUDA

gpu_sol = dislou_solve(
    cu(H), cu(ψ0), tlist, cu.(c_ops);
    e_ops=cu.([n_op]),
    gauge_set=sc,
    ntraj=ntraj,
    layer3_sizes=[20, 60],
    residual_tolerance=1e-4,
)
\end{lstlisting}
Both the spectral preparation and the propagation of each trajectory then run on the GPU, while \code{mcsolve} still runs the ensemble of trajectories on the CPU. In particular, the spectral preparation can be substantially accelerated in this way.

\section{Timing of the Kerr resonator transients}
\label{app:kerr_correctness_timings}

We complement the accuracy benchmarks of Fig.~\ref{fig:kerr_correctness} by timing the same short transients. For each trajectory ensemble, we compare the occupation $\hat n=\hat a^\dagger \hat a$ and quadrature $\hat x=(\hat a+\hat a^\dagger)/\sqrt{2}$ with the master-equation reference through
\begin{equation}
E=
\max_{\hat O\in\{\hat n,\hat x\}}
\frac{
\left[
N_t^{-1}\sum_{k=1}^{N_t}
\left(
\langle \hat O\rangle(t_k)-\langle \hat O\rangle_{\rm ME}(t_k)
\right)^2
\right]^{1/2}
}{
\max_k\langle \hat O\rangle_{\rm ME}(t_k)
-\min_k\langle \hat O\rangle_{\rm ME}(t_k)
},
\label{app}
\end{equation}
with $N_t=201$ equally spaced times over $0\leq\kappa t\leq5$. We denote by $\mathrm{UB}_{95}(E)$ the $97.5$th percentile of $500$ estimates obtained by resampling blocks of $50$ trajectories.

\begin{figure}[!htbp]
\centering
\reportfig{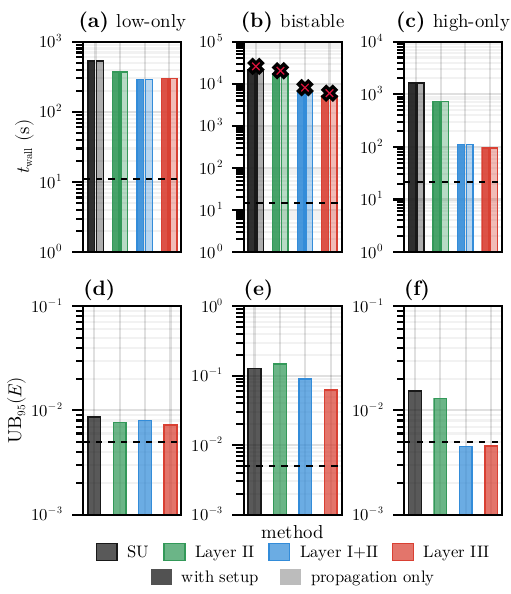}{\columnwidth}
\caption{Timing and accuracy for the short Kerr resonator transients of
Fig.~\ref{fig:kerr_correctness}. Columns show the dim monostable
(low-only), bistable, and bright monostable (high-only) regimes.
Top: wall time at the first sampled cumulative ensemble satisfying
$\mathrm{UB}{95}(E)\leq0.005$ for standard unraveling, Layer~II,
Layers~I+II, and Layer~III. Crosses indicate that the target accuracy was not
reached within $10^5$ trajectories. Dark bars include setup time,
while light bars show propagation time only. Bottom:
$\mathrm{UB}{95}(E)$ for the largest cumulative ensemble completed
within a $120$~s budget. Dashed lines mark the accuracy
target.}
\label{fig:kerr_correctness_timings}
\end{figure}

Figure~\ref{fig:kerr_correctness_timings} shows both the wall time required to reach $\mathrm{UB}_{95}(E)\leq0.005$ and the accuracy obtained within a fixed $120$~s  budget. Layers~I+II and III reach the fixed-accuracy target before standard unraveling in both monostable regimes, while no method reaches it within $10^5$ trajectories in the bistable case. At fixed wall time, both methods also yield a smaller $\mathrm{UB}{95}(E)$ than standard unraveling for all three transients.

\section{Scaling of the activity in the Kerr resonator as a function of photon number}
\label{app:kerr_activity_scaling}

The timing comparison of Sec.~\ref{sec:kerr_performance} combines two effects: the cost of propagating each no-jump interval and the number of such intervals generated by the selected unraveling. To isolate the latter, we vary the occupation while preserving the overall structure of the dynamics and measure how the locally optimal unraveling suppresses the number of quantum jumps.

We consider the Kerr resonator in the bright monostable regime and vary its steady-state occupation $\bar n$ from $5$ to $100$. Starting from a reference model with occupation $\bar n_{\rm ref}$, we define
$s=\bar n/\bar n_{\rm ref}$ and rescale
\begin{equation}
K=\frac{K_{\rm ref}}{s},
\qquad
\varepsilon=\varepsilon_{\rm ref}\sqrt{s},
\label{eq:app-kerr-scaling}
\end{equation}
while keeping $\Delta$, $\kappa$, and the propagation time $T$ fixed. This scaling keeps both $K\bar n$ and $\varepsilon/\sqrt{\bar n}$ constant, so that increasing $\bar n$ does not substantially alter the phase-space structure of the dynamics.

\begin{figure}
  \centering
  \reportfig{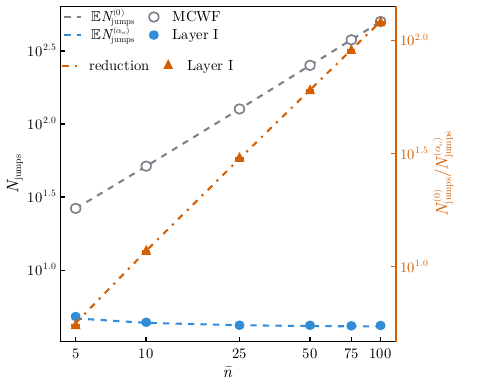}{\columnwidth}
  \caption{Jump counts as a function of the steady-state occupation $\bar n$ for the
Kerr resonator.
Gray open circles show the measured photon-loss counts in the undisplaced
unraveling, and the gray dashed curve gives the corresponding prediction
from Eq.~\eqref{eq:app-kerr-scaling-original}.
Blue filled circles and the blue dashed curve show the measured and
predicted counts for Layer~I with
$\zeta=-\sqrt{\kappa}\,\alpha_{\rm h}$.
Orange triangles (right axis) show the measured ratio between the
undisplaced and displaced counts, while the orange dash-dotted curve gives
the ratio of their predictions.
Each trajectory is initialized in a coherent state at the corresponding
semiclassical bright fixed point and evolved to $\kappa T=5$.}
  \label{fig:kerr_activity_scaling}
\end{figure}

For each value of $\bar n$, we initialize the trajectory in a coherent state centered on the corresponding stable semiclassical fixed point, whose amplitude we denote by $\alpha_{\rm h}(\bar n)$. Layer~I uses the associated gauge
\begin{equation}
\zeta=-\sqrt{\kappa}\alpha_{\rm h}(\bar n),
\end{equation}
which removes the coherent contribution to the photon-loss jump counts at the fixed point.

The expected number of jumps over the propagation window follows directly from the time-integrated activity of Eq.~\eqref{eq:physical_rate}. In the undisplaced and displaced gauges, respectively,
\begin{align}
\mathbb E N_{\rm jumps}^{(0)}
&=\kappa T
\overline{\langle \hat a^\dagger \hat a\rangle},
\label{eq:app-kerr-scaling-original}
\\
\mathbb E N_{\rm jumps}^{(\alpha_{\rm h})}
&=\kappa T
\overline{
\langle
(\hat a-\alpha_{\rm h})^\dagger
(\hat a-\alpha_{\rm h})
\rangle},
\label{eq:app-kerr-scaling-prediction}
\end{align}
where the overline denotes the time average over the interval $[0,T]$. The undisplaced activity contains the coherent photon-loss contribution and therefore grows with $\bar n$. By contrast, the displaced activity contains only the fluctuations about $\alpha_{\rm h}$ and the residual mismatch
$\lvert\langle \hat a\rangle-\alpha_{\rm h}\rvert^2$ appearing in Eq.~\eqref{eq:physical_rate}, neither of which grows with the coherent occupation under the scaling above.

Figure~\ref{fig:kerr_activity_scaling} compares these predictions with the jump counts measured along the corresponding trajectories. The undisplaced count increases strongly with $\bar n$, whereas the Layer~I count varies only weakly, in agreement with Eq.~\eqref{eq:app-kerr-scaling-prediction}. Consequently, the reduction in the number of jumps becomes increasingly pronounced as the resonator becomes brighter. As expected, the locally optimal unraveling therefore removes the occupation-dependent coherent contribution from the measurement record, rather than providing a fixed-factor reduction in activity.

Overall, these benchmarks verify the claims of Sec.~\ref{sec:method} in a model for which master-equation evolution remains feasible. The ensemble averages agree with the master-equation reference within sampling noise, while the propagation cost of trajectories initialized in the bright and dim regions is reduced by approximately $2\times10^2$ and $9\times10^4$, respectively. The occupation sweep further shows that this gain tracks the reduction in event activity predicted by Eq.~\eqref{eq:gauge_rate}.

\newpage
\bibliographystyle{quantum-noquotes}
\bibliography{refs}

\end{document}